\documentstyle[12pt,epsfig]{article}

\def\shiftdown#1{#1\llap{\lower.04ex\hbox{#1}}}

\begin{document}

\vspace*{0.3cm}

\begin{center}
{\bf {\large 
 Conformal group of transformations of the
quantum field operators in the momentum space
 and the five dimensional Lagrangian approach} }

\end{center}

\begin{center}
{\large {\it A.\ I.\ Machavariani$^{\diamond \ast}$ }}

{\em $^{\diamond }$ Joint\ Institute\ for\ Nuclear\ Research,\ Moscow
Region \\
141980 Dubna,\ Russia}\\[0pt]
{\em $^{*}$ High Energy Physics Institute of Tbilisi State University,
University Str. 9 }\\[0pt]
{\em 380086 Tbilisi, Georgia }
\end{center}

\vspace{0.5cm} \medskip

\begin{abstract}

Conformal group of
transformations in the momentum space,
 consisting of 
translations $p'_{\mu}=p_{\mu}+h_{\mu}$,  rotations
$p'_{\mu}=\Lambda^{\nu}_{\mu}p_{\nu}$, dilatation
$p'_{\mu}=\lambda p_{\mu}$ and inversion $p'_{\mu}=
-M^2p_{\mu}/p^2$ of the four-momentum $p_{\mu}$, is used for the  five
dimensional generalization of the
equations of motion for the interacting massive particles.
It is shown, that the ${\cal S}$-matrix of the 
charged and the neutral particles
scattering is invariant under  translations in a
 four-dimensional momentum space $p'_{\mu}=p_{\mu}+h_{\mu}$.
In the suggested system of equations of
motion, the one-dimensional equations over the fifth coordinate
$x_5$ are separated and these one dimensional equations have the
form of the evaluation equations with $x_5=\sqrt{x_o^2-{\bf
x}^2}$.

The important
property of the  derived five dimensional equations of motion
is the explicit separation of the parts of these equations
according to  the inversion
 $p'_{\mu}=-M^2 p_{\mu}/p^{2}$, where $M$
is a scale constant.

In the framework of the considered formulation the five
dimensional generalization of the nonlinear $\sigma$-model is
obtained. The appropriate five dimensional  Lagrangian coincides
with the usual nonlinear $\sigma$-model  in the region
$p^{\mu}p_{\mu}>M^2$. The scale
parameter $M$  is determined by the pion mass  $2M^2=
m_{\pi}^2$. Unlike  the usual nonlinear $\sigma$ model, in
the proposed Lagrangian the chiral symmetry breaking term
$\lambda\sigma$ is exactly reproduced.
 For the Lagrangian with the spontaneous broken
 $SU(2)\times U(1)$ symmetry the scale parameter
$M^2$ is determined by the Higgs particle mass
$8m^2_{Higgs}=9M^2$. In addition, the scalar particle interaction
Lagrangian has different signs in the regions
$p^{\mu}p_{\mu}>M^2$ and $0<p^{\mu}p_{\mu}<M^2$.

\end{abstract}

\eject

\vspace{0.5cm}

{\large \bf \ \ \ \ \ \ Content} \vspace{0.2 cm}

\begin{itemize}

\item[ ]
{\bf \ \ \ \ \ Introduction \hfill 3}

\item[\&1.]
{\bf  Conformal transformations in the 4D momentum space\\
and the scattering ${\cal S}$-matrix \hfill 4} \vspace{0.15cm}

\item[\&2.]
{\bf Five dimensional projection
\hfill 10}\vspace{0.15cm}

\item[\&3.]{\bf{
4D and 5D  equation of motion
in the coordinate space
\hfill 14}}
\vspace{0.15 cm}

\item[\&4.]
{\bf{Lagrangian approach
\hfill 18}}\vspace{0.15 cm}

\item[\&5.]
{\bf{ Construction of  5D Lagrangian
  ${\cal L}_{INT}(x,x_5)$ via  $l_a(x,x_5)$ (3.4)\hfill 21}}
\vspace{0.15 cm}

 {\em I.\ \ \ Off shell extension
\hfill 21}\vspace{0.15 cm}

{\em II.\ \ \ Nonlinear $\sigma$ model\hfill 24}

\item[\&6.]
{\bf{Models with the gauge transformations
\hfill 25}}
\vspace{0.15 cm}

 {\em I.\ \ \ Gauge transformations in the 4D and in the 5D spaces 
\hfill 25}\vspace{0.15 cm}

{\em II.\ \ \ Gauge $SU(2)\times U(1)$ theory\hfill 27}

\vspace{0.15 cm}
\item[\&7.]
{\bf Conclusion\hfill 27}

\end{itemize}

\newpage

\vspace{0.5cm}

\begin{center}
{\bf{ Introduction}}
\end{center}

\vspace{0.5cm}

Conformal transformations of the field operators 
and corresponding equation of motion 
in the momentum space were considered in ref.
\cite{Kas,Ca2,BR,Budinich}. In these papers  conformal
transformations were performed in the coordinate space 
and corresponding relations were constructed in the momentum space 
using the Fourier transform. Therefore in this approach 
the translation operator coincides with the four momentum
${\widehat P}_{\mu}=p_{\mu}$ in the momentum space. Two particular features 
determine the advantages of the conformal transformations in
the momentum space\cite{BR}. First, the real observables of the
particle interactions, like the cross sections and the corresponding
scattering amplitudes, are determined in the momentum space.
Secondly, the accuracy of the measurement of the particle coordinates
is in principle restricted by the Compton length of this
particle. Moreover in the conformal invariant case
 the determination of the coordinates
 of the massless particle 
generates an additional essential trouble 
(see  \cite{BaR} ch. 20 and \cite{Bacr}). 

In the present paper we consider conformal transformations  
of the off shell four momentum 
$q_{\mu}$ ($q_o\ne\sqrt{{\bf q}^2+m^2}$). In this case 
the generator
of the translation in the four-momentum space  
${\widehat P}_{\mu}=i\partial / \partial q_{\mu}$ 
 coincides with the corresponding coordinate ${\widehat
   P}_{\mu}=x_{\mu}$ in the coordinate space.
Translations, rotations, scale transformation and inversion 
of four momentum $q_{\mu}$  forms the conformal group
because the metric tensor of this space transforms under these
transformations  as 
 $g'_{\mu\nu}(q')=\bigl(1+f(q)\bigr)g_{\mu\nu}(q)$.

Conformal transformations and their 
numerous applications are presented in  
many books and review papers (see for instance 
\cite{Braun,Fradkin,Todorov,BK,Kon,Gatto,Alf,Ca1}).
The 6D representation of conformal transformations 
was done in the Dirac geometrical model \cite{Dir,M1,Gatto}, where
each of the conformal transformations was single-valued reproduced
via the appropriate 6D rotation in the invariance 6D cone
$\kappa_A\kappa^A\equiv\kappa_{\mu}\kappa^{\mu}+\kappa_5^2-\kappa_6^2=0$,
where the four momentum
$q_{\mu}$ ($\mu=0,1,2,3$) is defined as
$q_{\mu}=M \kappa_{\mu}/{(\kappa_5+\kappa_6)}$ and
$M$ is a scale parameter.
The invariance of the 6D cone
$\kappa_A\kappa^A=0$ is valid for  each of the conformal transformation
even when the conformal invariance is violated by the mass or other
dimensional parameters of the interacting particle.  
Therefore we use this invariance of 6D cone as a constraint for
the derivation of the equation of motion for arbitrary interacting 
massive field. In particular, we will project
$\kappa_A\kappa^A=0$  on the two 5D surface
$q_{\mu}q^{\mu} + q_5^2= M^2$ and
$q_{\mu}q^{\mu} - q_5^2= -M^2$, so that 
these 5D hyperboloids 
are  also invariant under the conformal transformations. 
This invariance enables us to introduce the constraints 
$\Bigl(q_{\mu}q^{\mu} \pm q_5^2 \mp M^2\Bigr)\Phi_{\pm}(q,q_5)=0$
for the 5D field operators.
Afterwards we  construct corresponding  5D Lagrangians and
the 5D equation of motion and consider their 4D reductions.

Invariance of  quantum field theory
under the four-momentum translation $q'_{\mu}=q_{\mu}+h_{\mu}$ 
in the homogeneous 4D momentum space is interpreted
as invariance under  a gauge transformation of the  charged field
$\Phi'_{\gamma}(x)=\Phi_{\gamma}^h(x)=e^{ih_{\mu}x^{\mu}}\Phi_{\gamma}(x)$.
We will show, that
for a neutral field the four-momentum translation
has  the more complicated
form $\Phi'_{\gamma}(x)=\Phi_{\gamma}^{h}(x)$ which
does not change the creation or annihilation operator
of the considered particle. Thereby the
scattering ${\cal S}$-matrix is invariant under translations in the
four-dimensional momentum space $q'_{\mu}=q_{\mu}+h_{\mu}$.
Moreover, one can generalize
invariance of the 6D form $\kappa_A\kappa^A=0$ and these 5D
 projections under the translations $q'_{\mu}=q_{\mu}+h_{\mu}$ 
for  other kinds gauge transformations
$q'_{\mu}=q_{\mu}-eA_{\mu}(q)$. 
This  enables us to extend
suggested 5D approach 
for the  models with the gauge transformations.

The 5D version of the quantum  field theory 
with the invariance form 
$q_{\mu}q^{\mu} + q_5^2= M^2$ or $q_{\mu}q^{\mu} - q_5^2= -M^2$ 
was suggested in refs.  \cite{K1,K2,K3},
where the scale  parameter $M$ was interpreted as the 
fundamental (maximal) mass
and its inverse $1/M$ as the fundamental (minimal) 
length \cite{Heisenberg,Markov}.
In the present formulation $M$ has the meaning of a boundary parameter
which may be determined 
in the theories with a spontaneous  symmetry breaking.
In particular, for the  nonlinear $\sigma$ model  $M$ is fixed via
the mass of $\pi$ meson and for the standard model with the
spontaneous
 $SU(2)\times U(1)$ symmetry  breaking the scale parameter
$M$ is determined by the  mass of the Higgs particle.

\begin{center}
{\bf{1. \ Conformal transformations 
 in the 4D momentum space and the scattering ${\cal S}$-matrix}}
\end{center}

\vspace{0.5cm}

Conformal transformations of a four-momentum
 $q_{\mu}$ $(\mu=0,1,2,3)$  compounded of the following   ¨
transformations:\newline
translations
$$T(h):\ \ \ \ \ \ \ \ \ \ \ \
\ \ q_{\mu}\longrightarrow q_{\mu}'=q_{\mu}+h_{\mu},\eqno(1.1a)$$
rotations

$$R(\Lambda):\ \ \ \ \ \ \ \ \ \ \ \ \ \ \ \ \ q_{\mu}\longrightarrow
q_{\mu}'=\Lambda_{\mu}^{\nu}q_{\nu}
,\eqno(1.1b)$$
dilatation
$${\cal D}(\lambda):\ \ \ \ \ \ \ \ \ \ \ \ \ \ \ \ \ \ \ \
q_{\mu}\longrightarrow q_{\mu}'=e^{\lambda} \
q_{\mu},\eqno(1.1c)$$ 
and inversions
$$I(M^2):\ \ \ \ \ \ \ \ \ \ \ \ \ \ \ \ \ \ \ \
q_{\mu}\longrightarrow
q_{\mu}'=-M^2 q_{\mu}/q^2,\eqno(1.1d)$$ 
where a scale parameter $M$ insures the correct dimension of  $q_{\mu}'$.
Translation $T({\hbar})$ and inversions $I(M^2)$ form the special conformal
transformations
$${\cal K}(M^2,{\hbar})\equiv I(M^2)T({\hbar})I(M^2):\ \ \ \ \
q_{\mu}\longrightarrow q_{\mu}'={{ q_{\mu}-{\hbar}_{\mu} q^2/ M^2} \over
{1-2q_{\nu}{\hbar}^{\nu}/M^2+{\hbar}^2q^2/M^4} }.\eqno(1.1e)$$
\par
Obviously, $q_{\mu}$ in (1.1a)-(1.1e) is off mass shell 
($q_o\ne\sqrt{{\bf q}^2+m^2}$). Hereafter the on mass shell 
4D momenta will be denoted as $p_{\mu}$ ($p^2=m^2$, $p_o\ge 0$).

Following the Dirac geometrical model \cite{Dir},
transformations (1.1a)-(1.1e) may by realized as rotations in the 
6D space with the metric
$g_{AB}=diag(+1,-1,-1,-1,+1,-1)$ and on the 6D cone

$$\kappa^2\equiv\kappa_A\kappa^A=
\kappa_{\mu}\kappa^{\mu}+\kappa_5^2-\kappa_6^2=0,\eqno(1.2)$$
where according to  conformal covariant formulation \cite{M1,M2}, 

$$q_{\mu}={ {\kappa_{\mu}}\over{\kappa_{+} }};\ \ \ \
\kappa_{+}=(\kappa_{5}+\kappa_{6})/M;\ \ \ \
\mu=0,1,2,3;\eqno(1.3)$$ 
where $\kappa_{+}$ is a dimensionless scale parameter and
$\kappa_{\mu}$, $q_{\mu}$ and $M$ have the same dimensions
 in the system of units $\hbar=c=1$.


\par
{\underline {\bf Conformal transformations (1.1a)-(1.1e)
of a particle field operator 
$\Phi_{\gamma}(x)$:}}

The particle field operator $\Phi_{\gamma}(x)$ 
with a spin-isospin quantum numbers $\gamma$ is usually expanded in the
positive and in the negative frequency parts in the 3D Fock space
$$\Phi_{\gamma}(x)=\int {{d^3 p}\over{(2\pi)^3 2\omega_{{\bf p}}} }
\Bigl[
a_{{\bf p}{\gamma}}(x_0)e^{-ipx}+{b^+}_{{\bf p}{\gamma}}(x_0)e^{ipx} \Bigr];
\ \ \ p_o\equiv \omega_{{\bf p}}
=\sqrt{ {\bf p}^2+m^2},\eqno(1.4a)$$
where in the asymptotic regions
$a_{{\bf p}{\gamma}}(x_0)$ and ${b^+}_{{\bf p}{\gamma}}(x_0)$ 
transforms into particle
 (antiparticle) annihilation (creation) operators
$lim_{x_0\to\pm \infty}<m|a_{\bf p\gamma}(x_0)|n>= <m|a_{\bf
p\gamma}(out,in)|n>$; ¨ $lim_{x_0\to\pm \infty}<m|{b^+}_{\bf
p\gamma}(x_0)|n>= <m|{b^+}_{\bf p\gamma}(out,in)|n>$, where
$n,m$ are  arbitrary asymptotic states.
On the other hand, $\Phi_{\gamma}(x)$  may be expanded in the 4D momentum 
space

$$\Phi_{\gamma}(x)=\int {{d^4 q}\over{(2\pi)^4  } }
\Bigl[{\Phi_{\gamma}^{(+)}} (q)e^{-iqx}+
{\Phi_{\gamma}^{(-)}}^+ (q)e^{iqx}\Bigr].\eqno(1.4b)$$

After comparison of the  expressions (1.4a) and (1.4b) we get

$${{e^{-i\omega_{\bf p}x_o}}\over{2\omega_{\bf p}}}
a_{\bf p\gamma}(x_o)=
\int {{dq_o}\over{2\pi}} {\Phi_{\gamma}^{(+)}} (q_o,{\bf p}) e^{-iq_ox_o}
\eqno(1.5a)$$
$$=i{{e^{-i\omega_{\bf p}x_o}}\over{2\omega_{\bf p}}}
\sum_{\beta}\int d^3x
<0|\Phi_{\beta}(x)|{\bf p}\gamma> \Bigl[{{\partial
\Phi_{\beta}(x)} \over{\partial x_o}}-i\omega_{\bf p}
\Phi_{\beta}(x)\Bigr]\eqno(1.5b)$$ and
$${{e^{i\omega_{\bf p_a}x_o}}\over{ 2\omega_{\bf p_a}} }
{b^+}_{\bf p_a\gamma}(x_0)=
\int {{dq_o}\over{2\pi}} {\Phi_{\gamma}^{(-)}}^+(q_o,{\bf p_a})e^{iq_ox_o}.
\eqno(1.6a)$$

$$=-i{{e^{i\omega_{\bf p_a}x_o}}\over{ 2\omega_{\bf p_a}} }
\sum_{\beta}\int d^3x
<{\bf p}_a\gamma|\Phi_{\beta}(x)|0> \Bigl[{{\partial
\Phi_{\beta}(x)} \over{\partial x_o}}+i\omega_{\bf p_a}
\Phi_{\beta}(x)\Bigr]\eqno(1.6b)$$
 where we have used the following expressions for a one-particle (antiparticle)
 states

$$<0|\Phi_{\beta}(x)|{\bf p}\gamma>=
Z^{-1/2}\int {{d^3p'}\over
{(2\pi)^32\omega_{\bf p'}}}e^{ip'x}<0|a_{{\bf p'}\beta}(x_o)|{\bf p}\gamma>
=Z^{-1/2}\delta_{\gamma\beta}e^{ipx},
\eqno(1.7a)$$

¨

$$<{\bf p}_a\gamma|\Phi_{\beta}(x)|0>=
Z_a^{-1/2}\int {{d^3p'}\over
{(2\pi)^32\omega_{\bf p'}} }e^{ip'x}
<{\bf p}_a\gamma|{b^+}_{{\bf p'}{\gamma}}(x_0)|0>
=Z_a^{-1/2}\delta_{\gamma\beta}e^{-ip_ax},
\eqno(1.7b)$$ where the index ${\sl a}$
 denotes the antiparticle state and
  $Z$ is the renormalization constant.

The field operators $a_{{\bf p}\gamma}(x_0)$ and ${b^+}_{{\bf p}{\gamma}}(x_o)$
are simply defined via the corresponding source operator
$\partial /
\partial{ x_{0}}a_{{\bf p}\beta}(x_0)=i\int d^3x e^{ipx}
j_{\beta}(x)$,
where
$\Bigl({{\partial^2} /{\partial{ x_{\mu}}\partial{x^{\mu} }
}}+m^2\Bigr) \Phi_{\beta}(x)=j_{\beta}(x)$. Moreover,
these operators determine the transition  ${\cal S}$-matrix 

$${\cal S}_{mn}\equiv
<out;{\bf p'}_{1} {\alpha'}_1,...,{\bf p'}_{m}{\alpha'}_m| {\bf
p}_{1} {\alpha}_1,...,{\bf p}_{n}{\alpha}_n;in>=
\prod_{i=1}^m\Bigl[ \int d{{x^0}'}_i {{d}\over{d {x^0}_i}}\Bigr]$$
$$\prod_{j=1}^n\Bigl[ \int d{x^0}_j {{d}\over{d {x^0}_j}}
\Bigr]<0|T\Bigl( a_{{\bf
p'}_m\alpha'_m}({x^0}'_m),...,a_{{\bf p'}_1\alpha'_1}({x^0}'_1)
a_{{\bf p}_n\alpha_n}^+({x^0}_n),...,a_{{\bf
p}_1\alpha_1}^+({x^0}_1)\Bigr)|0>. \eqno(1.8)$$

Next we will consider the transformations of $\Phi_{\beta}(x)$
 according to the conformal transformations of 
${\Phi_{\gamma}^{(\pm)}}(q)$  

$${\Phi_{\gamma}^{(\pm)}}(q)\to {\Phi_{\gamma}^{(\pm)}}'(q')
={\cal U}(g) {\Phi_{\gamma}^{(\pm)}}(q)  {{\cal U}}^{-1}(g) =
{\cal T}^{\beta}_{\gamma} \Phi_{\beta}^{(\pm)}(g^{-1}q),\eqno(1.9)$$ 
where $g$ indicates one of the (1.1a)-(1.1e) transformations 
$g\equiv\Bigl(T(h),R(\Lambda),{\cal D}(\lambda),{\cal K}(M,h)\Bigr)$,
${\cal T}^{\beta}_{\gamma}$ is the spin-isopin matrix and ${\cal U}(g)$
are defined through the generators of the corresponding transformations
in the well known form:

$$T(h):\ \ \ \ \ {\cal U}(h)=e^{ih_{\mu}{\cal X}^{\mu}};\ \ \
\ \ \biggl[{\cal X}_{\mu},{\Phi_{\gamma}^{(\pm)}} (q)\biggr]
=-i{{\partial} \over{\partial q^{\mu}} }{\Phi_{\gamma}^{(\pm)}}
(q), \eqno(1.10a)$$ ¯

$$R(\Lambda):\ \ \ \ \ {\cal U}(\Lambda)=
e^{i\Lambda_{\mu\nu}{\cal M}^{\mu\nu}};\ \ \ \biggl[{\cal
M}_{\mu\nu},{\Phi_{\gamma}^{(\pm)}} (q)\biggr]
=-i\biggl(q_{\mu}{{\partial} \over{\partial q^{\nu} } }\ - \
q_{\nu}{{\partial} \over{\partial q^{\mu} }
}-i\Sigma_{\mu\nu}\biggr) {\Phi^{(\pm)}}(q) \eqno(1.10b)$$ where
$\Sigma_{\mu\nu}=0$ for scalars,
$\Sigma_{\mu\nu}=i/4[\gamma_{\mu},\gamma_{\nu}]$ for fermions and
$\bigl(\Sigma_{\mu\nu}V_{\rho}\bigr)=ig_{\mu\rho}V_{\nu}-ig_{\nu\rho}V_{\mu}$
for the vectors $V_{\rho}$.

$${\cal D}(\lambda):\ \ \ \ \ {\cal U}(\lambda)=e^{i\lambda D};\ \ \
\ \ \biggl [D,{\Phi_{\gamma}^{(\pm)}} (q)\biggr]
=-i\biggl(q_{\mu}{{\partial} \over{\partial q^{\mu} }
}+id_m\biggr) {\Phi_{\gamma}^{(\pm)}} (q),\eqno(1.10c)$$ 
where $d_m$ indicates  the scale dimension of field. 
For example, in the scale-invariant case $d_m=-3$.
$${\cal K}(M,{\hbar}):\ \ \ \ \ {\cal U}({\hbar})=e^{i{\hbar}_{\mu}K^{\mu}};
\ \ \ \ \ \biggl[ K_{\mu},{\Phi_{\gamma}^{(\pm)}} (q)\biggr]
=-\biggl(2q_{\mu}D-q^2{\cal
X}_{\mu}+2iq^{\nu}\Sigma_{\mu\nu}\biggr) {\Phi_{\gamma}^{(\pm)}}
(q).\eqno(1.10d)$$

According to (1.4b)
the conformal transformations of the operators
 ${\Phi_{\gamma}^{(\pm)}}(q)$ (1.9) generates the corresponding 
transformations of $\Phi_{\gamma}'(x)$

$$\Phi_{\gamma}'(x)={\cal T}^{\beta}_{\gamma}
\int {{d^4 q}\over{(2\pi)^4  } }
\Bigl[{\Phi_{\beta}^{(+)}} (g^{-1}q)e^{-iqx}+
{\Phi_{\beta}^{(-)}}^+ (g^{-1}q)e^{iqx}\Bigr].\eqno(1.11)$$
In particular,  eq.(1.11) consists of the  following transformations:

\par
{\underline {\bf Four-momentum translation:}}

For a charged particle a four-momentum translation
is equivalent to  

$$q'_{\mu}=q_{\mu}+h_{\mu}\ \ \ \ \ \Longrightarrow
i{{\partial}\over{\partial x'_{\mu} }}=
i{{\partial}\over{\partial x_{\mu} }}+h_{\mu},
\eqno(1.12a)$$
which implies the well known gauge transformation of the charged particle
field operator  
$${\Phi_{\gamma}}'(x)=e^{ihx}\Phi_{\gamma}(x).\eqno(1.12b)$$

In order to get the  the gauge transformation 
formula (1.12b)  we introduce
 the following transformations
of ${\Phi_{\gamma}^{(\pm)}}(q)$

$${\Phi_{\gamma}^{(+)}}' (q)={\Phi_{\gamma}^{(+)}}(q+h);\ \ \
{{\Phi_{\gamma}^{(+)}}^+}' (q)={\Phi_{\gamma}^{(+)}}^+(q+h)\eqno(1.13a)$$
¨
$$ {\Phi_{\gamma}^{(-)}}' (q)={\Phi_{\gamma}^{(-)}}(q-h);\ \ \
{{\Phi_{\gamma}^{(-)}}^+}' (q)={\Phi_{\gamma}^{(-)}}^+(q-h).\eqno(1.13b)$$
After substitution of (1.13a,b) in (1.4b) we get

$${\Phi_{\gamma}}'(x)=\int {{d^4 q}\over{(2\pi)^4 } }
\Bigl[ {\Phi_{\gamma}^{(+)}}(q+h)e^{-iqx}+{\Phi_{\gamma}^{(-)}}^+(q-h)
e^{iqx}\Bigr]=e^{ihx}\Phi_{\gamma}(x)
\eqno(1.14).$$
In the same way we obtain

$$\int {{dq_o}\over{2\pi}} {\Phi_{\gamma}^{(+)}} (q_o+h_o,{\bf p+h}) 
e^{-iq_0x_0}=
\int {{dq_o}\over{2\pi}} {\Phi_{\gamma}^{(+)}} (q_o,{\bf p+h}) e^{-i(q_0-h_o)
x_0}$$

$$={{e^{-i\omega_{\bf p+h}x_o}}\over{2\omega_{\bf p+h}}}
a_{\bf p+h\gamma}(x_o) e^{ih_ox_o}
=Z^{1/2}
{{e^{i{\bf p+h}x}}\over{2\omega_{\bf p+h}}}
\sum_{\beta}
<0|\Phi'_{\gamma}(x)|{\bf p+h}\beta>
a_{\bf p+h\beta}'(x_o),
\eqno(1.15a)$$
where the operator
$$a_{\bf p\gamma}'(x_o)=i
 \sum_{\beta}\int d^3x <0|\Phi'_{\beta}(x)|{\bf p}\gamma>
{{ \stackrel{\longleftrightarrow}{\partial}}\over{\partial x^o}}
 \Phi'_{\beta}(x)\eqno(1.15b)$$
coincides with the operator (1.5b)
$$a_{\bf p\gamma}'(x_o)=a_{\bf p\gamma}(x_o).\eqno(1.15c)$$
Thus the gauge transformation (1.12a,b) generates
the following transformation of
the ${\cal S}$-matrix (1.8) 

$${\cal S'}_{mn}=
<out;{\bf p'_1+h}\ {\alpha'}_1,...,{\bf p'_m+h}\ {\alpha'}_m| {\bf
p_1+h}\ {\alpha}_1,...,{\bf p_n+h}\ {\alpha}_n;in>
\eqno(1.16)$$
This expression differs from ${\cal S}_{mn}$ by a shift of the position 
of the origin in the 3D momentum space. 
Therefore ${\cal S'}_{mn}={\cal S}_{mn}$
because only the relative momenta are physically
meaningful. 
A more complicated shift of a four-momentum operator 
${\widehat P}_{\mu}'={\widehat P}_{\mu}-<0|{\widehat P}_{\mu}|0>$
is often used in quantum field theory concerning the so-called 
zero-mode problem (see for example ch. 12 of \cite{BD}).

{\underline {\bf For\ a\ neutral\ particle }} field operator
$\phi_{\gamma}(x)$ the
 translation  $q_{\mu}'=q_{\mu}+h_{\mu}$ (1.12a) has a more complicated 
form due 
to absence of the antiparticle degree of freedom. In particular,
using (1.13a) we obtain 

$$\phi_{\gamma}'(x)=\int {{d^4 q}\over{(2\pi)^4 } } \Bigl[
{\phi_{\gamma}^{(+)}} (q)e^{-i(q-h)x}+ {\phi_{\gamma}^{(+)}}^+(q)
e^{i(q-h)x}\Bigr]\ne e^{ihx}\phi_{\gamma}(x).\eqno(1.17a)$$
or
$$\phi_{\gamma}'(x)=\int {{d^4 q}\over{(2\pi)^3}}
\delta((q_o-h_o)^2-({\bf q-h})^2-m^2)\theta(q_o-h_o)$$
$$\Bigl[
{\sl a}'_{{\bf q}{\gamma}}(x_0)e^{-i(\omega_{\bf q-h}-h_o)x_o+i({\bf q-h})
{\bf x} }+{{{\sl a}'}^+}_{{\bf q}{\gamma}}(x_o)
e^{i(\omega_{\bf q-h}-h_o)x_o+i({\bf q-h}){\bf x} }\Bigr],
\eqno(1.17b)$$

where

$${\sl a}_{\bf p\gamma}'(x_o)=i
 \sum_{\beta}\int d^3x <0|\phi'_{\beta}(x)|{\bf p}\gamma>
{{ \stackrel{\longleftrightarrow}{\partial}}\over{\partial x^0}}
 \phi'_{\beta}(x)\eqno(1.18)$$

According to (1.17 a,b) $\phi_{\gamma}'(x)$ remains  Hermitian
after translations
$q_{\mu}'=q_{\mu}+h_{\mu}$. On the other hand these
translations generate the nontrivial dependence of
$\phi_{\gamma}'(x)$ on  $h_{\mu}$.
The same dependence on the additional parameter $h_{\mu}$ 
appears in the real fields  $\phi_{1,2}(x)$
  constructed from the charged pion fields
 $\pi_{\pm}(x)$ after their gauge transformation (1.12a,b)
${\phi'}_{1}(x)=1/\sqrt{2}\Bigl(
\exp{(-ihx)}\pi_{+}(x)+\exp{(ihx)}{\pi_{+}}^+(x)
\Bigr)$ and

${\phi'}_{2}(x)=i/\sqrt{2}\Bigl(
\exp{(-ihx)}\pi_{+}(x)-\exp{(ihx)}{\pi_{+}}^+(x)
\Bigr)$.
It must be noted, that
a splitting of $\phi_{\gamma}(x)$ on the positive and the negative
frequency parts 
$\phi_{\gamma}(x)=\phi_{\gamma}^{(+)}(x)+{\phi_{\gamma}^{(+)}}^{\dagger }(x)$
can be realized 
with arbitrary
parameter $\alpha$ \cite{Weinberg}
as $\phi_{\gamma}(x)= e^{i\alpha}\phi_{\gamma}^{(+)}(x)+
e^{-i\alpha}{\phi_{\gamma}^{(+)}}^{\dagger }(x)$. 
In our case the additional dependence
of $\phi_{\gamma}´(x)$ on $h_{\mu}$ is result of the
condition (1.13a) which is necessary for the gauge
transformations rule (1.12a,b) of the charged field operators.

Using the ortho-normalization condition for functions
$f_{p-h}(x)=e^{i(p_o-h_o)x_o-i({\bf p-h}){\bf x} }$ we have

$$i\int f_{p'-h}^{\ast}(x)
{{\stackrel{\longleftrightarrow}{\partial} }\over{\partial x^0}}
f_{p-h}(x)d^3x=2(p_o-h_o)
(2\pi)^3\delta({\bf p'-p}),\eqno(1.19a)$$

$$
i\int f_{p'-h}(x)
{{\stackrel{\longleftrightarrow}{\partial}   }\over{\partial x^o}}
f_{p-h}(x)d^3x=i\int f_{p'-h}^{\ast}(x)
{{ \stackrel{\longleftrightarrow}{\partial}}\over{\partial x^0}}
f_{p-h}^{\ast}(x)d^3x=0.\eqno(1.19b)$$

It is easy to obtain
$${\sl a}_{\bf p\gamma}'(x_o)=a_{\bf p\gamma}(x_o), \eqno(1.20)$$
 where $a_{\bf p\gamma}(x_o)=i \sum_{\beta}\int d^3x
<0|\phi_{\beta}(x)|{\bf p}\gamma> {{
\stackrel{\longleftrightarrow}{\partial}}/{\partial x^0}}
 \phi_{\beta}(x)$. Relation (1.20) is analogue to the relation (1.15c)
for the charged fields.
This means, that ${\cal S}$-matrix 
transforms according to the same relation (1.16)
for the charged and neutral particles
 after translation in the 4D momentum space 
$q'_{\mu}=q_{\mu}+h_{\mu}$.
The dependence on the dummy variables $q_{o}$ and $q_{o}+h_{o}$ 
 disappears 
in the ${\cal S}$-matrix 
after the appropriate integration in eq.(1.5a), eq. (1.6a)
and in eq. (1.15a,b). 
 Thus for the ${\cal S}$-matrix and other observables the translation of 
$q\equiv(q_o,{\bf p})$ is reduced to the 3D translations
   ${\bf p'=p+h}$ which does not affect these observables.

{\underline{\bf Rotation (1.1b) and dilatation (1.1c) of $q_{\mu}$}}  
for the particle field operator $\Phi_{\gamma}(x)$ (1.11)
may be performed
using the rotations (1.10b)  and scale transformations
 (1.10c) of  ${\Phi_{\gamma}^{(\pm)}} (q)$ operators.
In particular, rotations $q'_{\mu}=\Lambda_{\mu\nu}q^{\nu}$
generates the following transformation of the field operators in the
configuration space
$$R(\Lambda):\ \ \ \ \
{\Phi_{\gamma}'}(x_{\mu})=\Phi_{\gamma}
(\Lambda_{\mu\nu}^{-1}x^{\nu}),\eqno(1.21)$$
and for dilatation $q_{\mu}'=e^{\lambda} \ q_{\mu}$   we have
$${\cal D}(\lambda):\ \ \ \ \
{\Phi_{\gamma}'}(x)=e^{4\lambda}\Phi_{\gamma}(e^{-\lambda}\ x).
\eqno(1.22)$$
Therefore the rotations and dilatation of $\Phi_{\gamma}(q)$
generate the analogical transformations of $\Phi_{\gamma}(x)$.

{\underline{\bf Special conformal transformation and inversion:}}
Special conformal transformation of  $q_{\mu}$ (1.1e) for  
 ${\Phi_{\gamma}^{(\pm)}}(q)$ has the form

$${\Phi_{\gamma}^{(+)}}'(q)={\Phi_{\gamma}^{(+)}}\Bigl((q^I+h)^I\Bigr);
\ \ \
{{\Phi_{\gamma}^{(+)}}^+}'(q)={\Phi_{\gamma}^{(+)}}^+\Bigl((q^I+h)^I\Bigr)
\eqno(1.23a)$$
¨
$$ {\Phi_{\gamma}^{(-)}}' (q)={\Phi_{\gamma}^{(-)}}\Bigl((q^I-h)^I\Bigr);\ \ \
{{\Phi_{\gamma}^{(-)}}^+}' (q)={\Phi_{\gamma}^{(-)}}^+\Bigl((q^I-h)^I\Bigr),
\eqno(1.23b)$$

where the index $^I$ relates to the inversion of $q_{\mu}$. 
According to (1.11) we get

$$\Phi_{\gamma}'(x)=
\int {{d^4 q}\over{(2\pi)^4 } }
\Bigl[{\Phi_{\gamma}^{(+)}}\Bigl((q^I+h)^I\Bigr)e^{-iqx}+
{\Phi_{\gamma}^{(-)}}^+\Bigl((q^I-h)^I\Bigr)e^{iqx}\Bigr].
\eqno(1.24)$$

This formula   can be essentially simplified if 
$\Phi_{\gamma}(x)$ is inversion invariant 

$$\Phi_{\gamma}^I(x)=
\int {{d^4 q}\over{(2\pi)^4 } }
\Bigl[{\Phi_{\gamma}^{(+)}}(q^I)e^{-iqx}+
{\Phi_{\gamma}^{(-)}}^+(q^I)e^{iqx}\Bigr]=\Phi_{\gamma}(x).\eqno(1.25)$$

Then after redefinition of the variables in (1.24)
${\Phi_{\gamma}^{(\pm)}}\Bigl((q^I\pm h)^I\Bigr)e^{\mp iqx}
\Rightarrow$
${\Phi_{\gamma}^{(\pm)}}\Bigl((q\pm h)^I\Bigr)e^{\mp iqx}\Rightarrow
{\Phi_{\gamma}^{(\pm)}}(q\pm h)e^{\mp iqx}$  we obtain the analogue to
 (1.12b) or (1.17) gauge transformation for $\Phi_{\gamma}'(x)$.

Arbitrary operator ${\Phi_{\gamma}^{(\pm)}}(q)$ 
may be divided into 
two parts

$${\Phi_{\gamma}^{(\pm)}}_{inv.}(q)={1\over 2}
\Bigl[{\Phi_{\gamma}^{(\pm)}}(q)+
{\Phi_{\gamma}^{(\pm)}}(q^I) \Bigr]\eqno(1.26a)$$

and

$${\Phi_{\gamma}^{(\pm)}}_{ps.-inv.}(q)={1\over 2}
\Bigl[{\Phi_{\gamma}^{(\pm)}}(q)-
{\Phi_{\gamma}^{(\pm)}}(q^I) \Bigr],\eqno(1.26b)$$
where ${\Phi_{\gamma}^{(\pm)}}_{inv.}(q)$ and 
${\Phi_{\gamma}^{(\pm)}}_{ps.-inv.}(q)$ denotes the inversion invariant 
and the inversion pseudo-invariant parts of ${\Phi_{\gamma}^{(\pm)}}(q)$. 

  The inversion invariant part of the complete field operator 
satisfies condition (1.25) for 
${\Phi_{\gamma}^{(\pm)}}_{inv.}(q)$. An analogous condition is valid
also for ${\Phi_{\gamma}^{(\pm)}}_{ps.-inv.}(q)$.
Expressions (1.26a,b)
  enable us to simplify
eq.(1.24) for the special conformal transformation of
${\Phi_{\gamma}^{(\pm)}}(x)$.

\vspace{0.7cm}

\centerline{\bf{2.\  Five dimensional projection}}

\vspace{0.4cm}

\par
The invariant form of the $O(2,4)$ group 
$\kappa_A\kappa^A=0$ (1.2) can be represented in the five dimensional form
with  $q_{\mu}$ (1.3) variables

$$q_{\mu}q^{\mu}+M^2{{\kappa_{-}}\over{\kappa_{+}}}
=0,\eqno(2.1a)$$
where
$$q_{\mu}={{\kappa_{\mu}}\over{\kappa_+}};\ \ \ \
\kappa_{\pm}={{ \kappa_5\pm\kappa_6}\over {M}}. \eqno(2.1b)$$
It is convenient to replace two variables
 $\kappa_{\pm}$ (or
$\kappa_5$, $\kappa_6$) in (2.1a) with one variable.
This procedure implies  a projection
of the six dimensional invariant cone
$\kappa_A\kappa^A=0$ into 5D space.  
It exists only two
 5D  De Sitter spaces with the following invariant forms of the 
$O(2,3)$ and $O(1,4)$ rotational groups \cite{BK,Todorov,K2} 

$$q_{\mu}q^{\mu}+q_5^2=M^2 \ \ \ \ \ \ \ \ \ \ \ \ \ \ \ \ \ \ \ \ \ \
 \ \ q_5^2 =M^2 { {2\kappa_{5}} \over {\kappa_{5}+\kappa_{6} } },\eqno(2.2a)$$

and

$$q_{\mu}q^{\mu}-q_5^2=-M^2 \ \ \ \ \ \ \ \ \ \ \ \ \ \ \ \ \ \ \ \ \ \ \
q_5^2 =M^2 { {2\kappa_{6}}\over {\kappa_{5}+\kappa_{6} } }.\eqno(2.2b)$$


\par
\footnotetext{ In the literature often is considered the stereographic  
projection of the 6D cone $\xi_A\xi^A=0$  into 4D
Minkowski space  with coordinates
$x_{\mu}=\xi_{\mu}{\ell}/(\xi_5+\xi_6)$, where
at the intermediate stage  are used
projections on the 5D hyperboloid
 $\eta_{\mu}\eta^{\mu}-\eta_5^2=- {\ell}^2$ (see for example
ch. 13 of \cite{IZ})  with  
 $\eta_{\mu}=\xi_{\mu}{\ell}/\xi_5$; $\eta_{5}=\xi_{6}{\ell}/\xi_5$ 
and
 $\eta_{\mu}=2x_{\mu}/(1-x^2/{\ell}^2);\ \ \ \eta_{5}={\ell}
 (1+x^2/{\ell}^2)/(1-x^2/{\ell}^2)$.
Here
$x^2=\ell^2(\eta_5/\ell-1)/(\eta_5/\ell+1)$ and at first sight
$x_{\mu}$ is  not restricted
by the 5D condition  $\eta_{\mu}\eta^{\mu}-\eta_5^2=- {\ell}^2$
like $q^2$  (2.2a,b)  in table  1 or 2.
Nevertheless, the 6D invariant form can be rewritten as
$x^{\mu}x_{\mu}+\ell^2(\xi_5-\xi_6)/(\xi_5+\xi_6)=0$ and the
appropriate projection  into  5D hyperboloid
$x^{\mu}x_{\mu}\pm x_5^2=\pm\ell^2$
with $x_5^2=2\xi_5(\ or\ \xi_6)\ell^2/(\xi_5+\xi_6)$
generates the corresponding restrictions.}

 $q_{\mu}$ and $q_5$ are real variable and they   
 are defined in the regions
$(-\infty,+\infty)$ and  $[0,+\infty)$ respectively\footnotemark.
In the considered formulation
 $q_{\mu}$ and $q_5$ are disposed in the hyperboloids
(2.2a,b). In particular, we place $0<q^2\le M^2$ 
in the hyperboloid  (2.2a) 
and $q^2>  M^2$ can be placed only
in the hyperboloid  (2.2b). Thus 
the conformal transformations for the whole $q^2$ values may be performed
 using  both hyperboloid  (2.2a,b). 
The values of $q_{\mu}$ and $q_5$ in these hyperboloids 
are singlevalued connected with each other
via inversion 
$q_{\mu}'=-M^2 q_{\mu}/q^2$ (1.1d). On the 6D cone 
$\kappa_A\kappa^A=0$ (1.2) inversion (1.1d) 
 can be carried out using the reflection of the    
$\kappa_6$ variable

$$I(M^2):\ \ \ \ \ \kappa_5^I=\kappa_5,\ \ \ \kappa_6^I=-\kappa_6;\ \ \
 ¨«¨\ \ \
\kappa_+^I=\kappa_-,\ \ \ \kappa_-^I=\kappa_+,\eqno(2.3)$$ 
which generates $q_{\mu}^I=-M^2 q_{\mu}/q^2$ according to
(2.1a,b). The advantage of  the 6D representation (2.3)
of the 4D transformation  $q_{\mu}^I=-M^2 q_{\mu}/q^2$ is that it
determines the transparent realization of the nonlinear 4D
transformation using the simple reflection in 6D space. 
In particular, for $0<q_{\mu}\le M^2$ on the hyperboloid
  $q^2+q_5^2=M^2$ , we have
${q^2}^I=M^4/q^2\ge M^2$ and
 $(-{q_5^2}+M^2)^I=
M^2 {\kappa_-}^I/{\kappa_+}^I =M^2/(\kappa_-/\kappa_+)
=M^4/({q_5}^2-M^2)$. Therefore, if $q_{\mu}^I$ belongs to  
(2.2b) hyperboloid ${q^2}^I+(-q_5^2+M^2)^I=0$, then
$q_{\mu}$ will be placed on the other hyperboloid because
${q^2}^I+(-q_5^2+M^2)^I=
M^4/[q^2(q_5^2-M^2)](q^2+q_5^2-M^2)=0$. 
The distribution of the  regions of the 
5D hyperboloid $q^2 \pm q_5^2=\pm M^2$
(2.2a,b) which covers the whole values of $q_{\mu}$ and $q_5$
 ($-\infty < q^2 < \infty$ and $q_5^2\ge 0$) 
 is given in table 1.

\par
\footnotetext{ The border point $q^2=0$ with
$q_5^2=M^2$ is included in the
domain $q_{\mu}q^{\mu}+
q_5^2=M^2$, because it belongs to the physical spectrum of the
massless particles.
After inversion $q^2=0$ 
transforms into $q^2=\infty$ of the hyperboloid  $q_{\mu}q^{\mu} -
q_5^2=-M^2$  in the region $II$.}

\vspace{0.7cm}

\centerline{\bf Table  1}
\vspace{0.2cm}

\hspace{-0.75cm}
\begin{tabular}{|c|c|c|c|c|} \hline\hline
        &        {\bf I}            &         {\bf II}
            &      {\bf III}            &         {\bf IV }
\\    \hline
           & $q^2+q_5^2=M^2$ & $q^2-q_5^2=-M^2$ & $q^2+q_5^2=M^2$
& $q^2-q_5^2=-M^2$ \\ \hline
 $ q^2$     &   $0\le q^2 \le M^2$   & $M^2< q^2< \infty$ &$-\infty<q^2<-M^2$
& $-M^2\le q^2< 0 $\\  \hline
 $ q_5^2$       &   $0\le q_5^2 \le M^2$   & $2M^2\le q_5^2< \infty$ 
                & $2M^2< q_5^2< \infty$ &
 $0\le q_5^2< M^2$ \\ \hline \hline
\end{tabular}
\vspace{0.5cm}

\par
In the region $I$ momenta   $q_{\mu}$
 are singlevalued connected 
with  $q_{\mu}$  in the region   $II$ 
  via inversion 
$\Bigl\{ {q_{\mu}} \Bigr\}_{II\ region}
=-M^2\Bigl\{q_{\mu}/q^2 \Bigr\}_{I\ region}$ 
and vice versa
$\Bigl\{ {q_{\mu}} \Bigr\}_{I\ region}
=-M^2\Bigl\{q_{\mu}/q^2 \Bigr\}_{II\ region}$.
In the same way are connected the four momenta in the regions 
$III$ and $IV$, where  $q^2<0$.
For $M\to\infty$ 5D spaces transforms into ordinary Minkowski space
with the domains $I$ and $IV$.
For  $M\to 0$  the regions $II$ and $III$ are remained.

\par
The scale transformation have the different form
 in the different areas in table 1.
In the 6D space the scale transformation $q_{\mu}'=e^{\lambda}q_{\mu}$ (1.1c)
 implies the rotation in the
 (6,5) plane. For $q^2\ge 0$ rotation
 $\kappa_5=M\ sh(\lambda);\kappa_6=M\ ch(\lambda)$
generates the following transformation of
$q^2$:\ \ \ $  q_{\mu}q^{\mu}=-
M^2{({\kappa_5-\kappa_6})/({\kappa_5+\kappa_6})
}=M^2e^{-2\lambda}$. For negative $q^2<0$  
we take $\kappa_5= M\ ch(\lambda);\
\kappa_6=M\ sh(\lambda)$  (i.e. $\kappa_+=e^{\lambda}$)
which gives
$q_{\mu}q^{\mu}=-M^2e^{2\lambda}$. 
The corresponding transformation of
$q_5^2$ with the related  $\lambda$, is given in table 2.


\newpage
\centerline{\bf Table 2}
\vspace{0.2cm}

\hspace{-1.5cm}
\begin{tabular}{|c|c|c|c|c|} \hline\hline
 {\rm rotation}          & \multicolumn{2}{|c}
{$\kappa_5=M\ sh\lambda;\  \kappa_6=M\ ch\lambda$}
& \multicolumn{2}{|c|}{$\kappa_5=M\ ch\lambda; \ \kappa_6=M\ sh\lambda$} \\
\hline
$\lambda$           &  $\lambda>0$ &  $\lambda<0$ & $\lambda<0$ &
$\lambda>0$  \\ \hline
        &        {\bf I}            &         {\bf II}
            &      {\bf III}            &         {\bf IV }
\\    \hline
{\rm hyperboloid} & $q^2+q_5^2=M^2$ & $q^2-q_5^2=-M^2$ & $q^2+q_5^2=M^2$
& $q^2-q_5^2=-M^2$ \\ \hline
 $ q^2$     &   $q^2=M^2e^{-2\lambda}$   & $q^2=M^2e^{-2\lambda}$ &
$q^2=-M^2e^{-2\lambda}$
& $q^2=-M^2e^{-2\lambda}$\\  \hline
 $ q_5^2$       &   $q_5^2=M^2(1-e^{-2\lambda})$   &
$q_5^2=M^2(1+e^{-2\lambda})$ & $q_5^2=M^2(1+e^{-2\lambda})$ &
 $q_5^2=M^2(1-e^{-2\lambda})$ \\ \hline \hline
\end{tabular}
\vspace{0.5cm}

\par
In table 2 it is shown, that the scale transformation 
parameter $\lambda$ (or $\kappa_+=e^{-\lambda}$) single-valued determines
$q_5^2$ and $q^2$.
In particular, in the region I
for hyperboloid $q^2+q_5^2=M^2$
the scale transformation  is realizable with $\lambda>0$,
which implies the compression of $q^2$ or  $q^2/M^2$.
In opposite to this,  in the region II in the hyperboloid  $q^2-q_5^2=-M^2$ 
  $\lambda<0$, i.e.
the same scale transformation generates the stretching
of $q^2$ or $q^2/M^2$.  
An analogical scale transformation can be observed for $q^2<0$, where
in the region III dilatation generates stretching and in the region IV we
have compression. 
In other words, 
projections of the 5D cone  $\kappa_A \kappa^A=0$
on the 5D hyperboloid for $q^2>0$ implies
$$\kappa_A \kappa^A=0\Longrightarrow
q_{\mu}q^{\mu}+q_5^2=M^2,\ \ \ \  for  \ \ 0\le q_{\mu}q^{\mu}\le M^2\ \
with\ \ \ \lambda\geq 0\eqno(2.4a)$$
$$\kappa_A \kappa^A=0\Longrightarrow
q_{\mu}q^{\mu}-q_5^2=-M^2,\ \ \ \  for  \ \ q_{\mu}q^{\mu}> M^2\ \ 
with\ \ \ \lambda\le 0.\eqno(2.4b)$$

Inversion ${q'}^2=M^4/q^2$  replaces
the internal points $0\le q^2<M^2$ (section I)
 by the external points $q^2>M^2$ (section II)
and vice versa. Therefore the region $I$  
my be defined as the internal region  ($0\le q_{\mu}q^{\mu}\le M^2$) and
the region $II$   ($ q_{\mu}q^{\mu}> M^2$)
as the external region. Thereby
in order to simplify the following notation  
the hyperboloid  $q^2+q_5^2=M^2$  we will  denote 
as the ``internal''
surface and the
hyperboloid  $q^2+q_5^2=M^2$  we will call as the 
``external'' ´domain.

\par
By translation $q_{\mu}'=q_{\mu}+h_{\mu}$ the 6D cone
 $\kappa_A\kappa^A=0$, as well as the 5D forms
(2.2a,b) $q^2\pm q_5^2=\pm M^2$ are preserved.
In particular, 
after the appropriate 6D rotations 
$\kappa_{\mu}'=\kappa_{\mu}+h_{\mu}\kappa_{+};$
$\kappa_{+}'=\kappa_{+};$ $\kappa_{-}'=\kappa_{-}+2/M^2
h_{\nu}\kappa^{\nu}+h^2/M^2\kappa_+$ we get
$${q^2}'=q^2+2 h_{\nu}q^{\nu}+h^2=-M^2{{\kappa_{-}'}\over{\kappa_{+}}};$$
$${q_5^2}'=q_5^2\pm(2 h_{\nu}q^{\nu}+h^2),\eqno(2.5)$$
where the sign $-$ corresponds to
  $q^2+q_5^2=M^2$ and $+$ relates
to  $q^2-q_5^2=-M^2$. Using (2.5) we get 
${q^2}'\pm{q_5^2}'={q^2}\pm{q_5^2}$. 
Nevertheless, the transformation
$q_{\mu}'=q_{\mu}+h_{\mu}$ can generate a transition from the
time-like region
$q^2\ge 0$ into space-like region $q^2<0$ i.e. transition
from the region $I\ or\ II$ into regions $III\ or\ IV$ correspondingly.
Transition between the $q^2> 0$ and $q^2< 0$ regions
is result of the transposition of  $\kappa^{6}$ and
$\kappa^{5}$ variables and we have used this transposition
 in table 2 for the scale transformations.

\par
It must be noted, that inversion transforms the generators of the
conformal group  (1.10a) - (1.10d)
in the following way $\  ^{\dagger}$\footnotemark
$${\cal X}_{\mu}=I(M^2)\ K_{\mu}\ I(M^2);\ \ \
{\cal M}_{\mu\nu}=I(M^2)\ {\cal M}_{\mu\nu}\ I(M^2);$$
$$\ \ \ D'=I(M^2)\ D\ I(M^2)=-D;\ \ \ K_{\mu}=I(M^2)\ {\cal X}_{\mu}\ I(M^2),
\eqno(2.6)$$

Therefore one can perform the conformal transformations 
only in the ``internal'' regions $I$ and $III$ and
 obtain the corresponding transformations in
the ``external'' regions $II$ and $IV$ using the inversion.

\par
\footnotetext{
An analogical transformation  can be performed using
the Weyl reflection, i.e. rotation through $90^o$ in the $(0,5)$
plane \cite{BR}.}

\vspace{0.5cm}

{\underline {\bf 5D reduction of the field operators:}}\ \ \ \ \
Next we have to connect a 6D field operator
$\phi^{(\pm)}(\kappa_A)$ defined in the cone $\kappa_A\kappa^A=0$
 ($A=\mu;5,6\equiv$ 0, 1, 2, 3; 5,6)
with the 5D operators
${\phi^{(\pm)}_{inr}}(q,q_5)$  and
${\phi^{(\pm)}_{ext}}(q,q_5)$, defined
in the surfaces $q^2\pm q_5^2=\pm M^2$ (2.2a,b).
Here index $\  ^{(\pm)}$ corresponds to positive or negative frequency,
the subscripts  $inr$ or
$ext$ indicate the surfaces   $q^2+q_5^2=M^2$ and $q^2+q_5^2=-M^2$
correspondingly.
In particular the internal (inr) area 
relates to the  $I$ and $III$ regions in  table 1
and the external (ext) regions corresponds to the sections
 $II$ and $IV$   in
table 1. Afterwards for the sake of simplicity we  omit the
spin-isospin indices ${\gamma}$.

\par
 According to the manifestly covariant construction of the  $O(2,4)$ 
conformal group \cite{M1,M2}, conformal transformations (1.1a)-(1.1e) 
are equivalent to the 6D rotations in the cone $\kappa_A\kappa^A=0$
with the following choice of the six independent variables

$$q_{\mu}=\kappa_{\mu}/\kappa_{+};\ \ \ \kappa_{+}=(\kappa_{5}+\kappa_{6})/M;
\ \ \ \kappa^2=\kappa^A \kappa_A.\eqno(2.7)$$
Only this choice of the  variables
makes  independent the generators of the conformal group $O(2,4)$
on  $\partial/\partial{\kappa^2}$.
In particular, for a spinless particle these generators are
$$ {\cal X}_{\mu}=i{{\partial}\over{\partial q^{\mu}}};\ \ \
{\cal M}_{\mu\nu}=i\Bigl(q_{\mu}{{\partial}\over{\partial q^{\nu}}}-
q_{\nu}{{\partial}\over{\partial q^{\mu}}}\Bigr);$$

$$D=i\Bigl(q_{\mu}{{\partial}\over{\partial
q^{\mu}}}-k_{+}{{\partial}\over{\partial k_{+}}}\Bigr);\ \ \
{\cal K}_{\mu}=2q_{\mu}D-q^2{\cal X}_{\mu}.\eqno(2.8)$$

Using the variables (2.7) the 6D field operator takes the form
$$\phi^{(\pm)}(\kappa_A)\equiv\phi^{(\pm)}(q_{\mu},\kappa_+,\kappa^2).
\eqno(2.9)$$
The homogeneous over the scale variable $\kappa_+$ operator
$\phi^{(\pm)}(q_{\mu},\kappa_+,\kappa^2)$  may be rewritten as
$$\phi^{(\pm)}(q_{\mu},\kappa_+,\kappa^2)= (\kappa_+)^d
\varphi^{(\pm)}(q_{\mu},\kappa^2),\eqno(2.10)$$
and for the 4D physical field operator. in analogue 
to \cite{M1,M2} we get
$$\Phi^{(\pm)}(q_{\mu})= (\kappa_+)^{-d}{\cal O}
\phi^{(\pm)}(\kappa_A),\eqno(2.11)$$
where $d$ defines the scale dimension of the considered operator,
and ${\cal O}$ 
acts on the spin-isospin variables.

\par
In the present paper we will use other  recipe
of projection of the 6D cone $\kappa^2=0$ 
 into 5D surfaces
$q_{\mu}q^{\mu}\pm q_5^2=\pm M^2$ and in the 4D momentum space.
In particular, we will treat the condition  $\kappa^2=0$ as the
dynamical restriction, i. e. we will require the validity of the
following constraint
$$\Bigl(\kappa^A\kappa_A\Bigr)\phi^{(\pm)}(q_{\mu},\kappa_+,\kappa^2)=
\kappa_{+}^2\Bigl(q_{\mu}q^{\mu}+M^2{{\kappa_{-}}\over{\kappa_{+}}}\Bigr)
\phi^{(\pm)}(q_{\mu},\kappa_+,\kappa^2)=0,
\eqno(2.12)$$
Projection of this equation on the 5D surfaces
 $q_{\mu}q^{\mu}\pm q^2_5=\pm M^2$ gives

$$\Bigl(q_{\mu}q^{\mu}+q_5^2-M^2\Bigr)\phi^{(\pm)}_{inr}
\Bigl(q_{\mu},\kappa_+,{\kappa_+}^2(q_{\mu}q^{\mu}+q_5^2-M^2)
\Bigr)=0,\eqno(2.13a)$$

$$\Bigl(q_{\mu}q^{\mu}-q_5^2+M^2\Bigr)\phi^{(\pm)}_{ext}
\Bigl(q_{\mu},\kappa_+,{\kappa_+}^2(q_{\mu}q^{\mu}-q_5^2+M^2)
\Bigr)=0.\eqno(2.13b)$$

As it was show in table 2, the magnitude of the scale parameter
$\kappa_+= e^{\lambda}$ is unambiguously defined via
$q^2$ or $q_5^2$ variables. Therefore we can introduce the 
5D fields
$$\varphi^{(\pm)}_{inr}(q_{\mu},q_5)\equiv (\kappa_+)^{-d}{\cal O}
\phi^{(\pm)}_{inr}\Bigl(q_{\mu},\kappa_+,{\kappa_+}^2
(q_{\mu}q^{\mu}+q_5^2-M^2)\Bigr),\eqno(2.14a)$$

$$\varphi^{(\pm)}_{ext}(q_{\mu},q_5)\equiv (\kappa_+)^{-d}{\cal O}
\phi^{(\pm)}_{ext}\Bigl(q_{\mu},\kappa_+,{\kappa_+}^2
(q_{\mu}q^{\mu}-q_5^2+M^2)\Bigr).\eqno(2.14b)$$

Then using  eq.(2.13a,b) we get 

$$\Bigl(q_{\mu}q^{\mu}+q_5^2-M^2\Bigr)\varphi^{(\pm)}_{inr}(q_{\mu},q_5)=0;
\eqno(2.15a)$$

$$\Bigl(q_{\mu}q^{\mu}-q_5^2+M^2\Bigr)\varphi^{(\pm)}_{ext}(q_{\mu},q_5)=0;
.\eqno(2.15b)$$

Equations (2.15a,b) present the desired 5D projections of the 6D
constraint (2.12). These relations can be treated also as the 4D equations,
because the fifth momentum on 
 $q^2 \pm q_5^2=\pm M^2$ shell is given 
$q_5=\pm \sqrt{|M^2 \mp q^2|}$.

\vspace{0.5cm}
\begin{center}

{\bf{3.\ 4D and 5D  equations of motion
 in the coordinate space.}}
\end{center}
\vspace{0.4cm}
\par
It is well known that the theories with any dimensional 
parameters are conformal non-invariant. 
Nevertheless the conformal transformations 
in momentum space in such theories are realizable
and these transformations require the invariance of 
the 6D form $\kappa^2\phi(\kappa)=0$ (2.12). 
Therefore we can consider this condition
and the appropriate 5D projections 
$\Bigl(q_{\mu}q^{\mu}\pm{q_5}^2\mp M^2\Bigr)
\varphi^{(\pm)}_{inr,ext}(q_{\mu},q_5)=0$ (2.15a,b)
as restrictions which can be  taken into account 
in the 4D equation of motion 

$$\biggl({{\partial^2}\over{\partial x^{\mu}\partial x_{\mu}}}+
m^2\biggr)\Phi(x)=J(x),\eqno(3.1a)$$
where
$$ J(x)=1/(2\pi)^{4}\int d^4q
\Bigl[e^{-iqx}J^{(+)}(q)+e^{iqx}J^{(-)}(q)\Bigr],\eqno(3.1b)$$ 
or
$$J(x)=\int {{d^3 p}\over{(2\pi)^3 2\omega_{{\bf p}}} }
\Bigl[{{\partial}\over{\partial x_0}} a_{{\bf
p}{\gamma}}(x_0)e^{-ipx}+ {{\partial}\over{\partial
x_0}}{b^+}_{{\bf p}{\gamma}}(x_0)e^{ipx} \Bigr]; \ \ \ p_o\equiv
\omega_{{\bf p}} =\sqrt{ {\bf p}^2+m^2}.\eqno(3.1c)$$

\par
In order to determine the relation between 4D eq.(3.1a) 
and 5D conditions  (2.15a,b) 
we introduce the following boundary conditions over the  fifth
coordinate $x_5$
$$\Phi(x)=\Phi(x,x_5=t_5)=\varphi_{inr}(x,x_5=t_5)+\varphi_{ext}(x,x_5=t_5)
\eqno(3.2a)$$ ¨
$${i\over M}{ {\partial}\over { \partial x_5}}\Phi(x,x_5)|_{x_5=t_5}
=\sum_{a=1,2}{i\over M}{ {\partial}\over { \partial
x_5}}\varphi_{a}(x,x_5) |_{x_5=t_5},\ \ \ where \ \ a=1,2\equiv inr,ext 
\eqno(3.2b)$$
and  $t_5$ is the same boundary value of $x_5$ which is convenient
to choose as $t_5=\tau=\sqrt{x_o^2-{\bf x}^2}$ or $t_5=0$, $a$
indicates  $inr$ or $ext$ operators.

Using (2.15a,b) we get 
$$\biggl({{\partial^2}\over{\partial x^{\mu}\partial x_{\mu}}}+
{{\partial^2}\over{\partial x^5\partial
x_5}}+M^2\biggr)\varphi_{inr}(x,x_5)=0 \eqno(3.3a)$$
 for the internal  hyperboloid with
 the regions $I,III$ in table 1 and ¨
$$\biggl({{\partial^2}\over{\partial x^{\mu}\partial x_{\mu}}}-
{{\partial^2}\over{\partial x^5\partial
x_5}}-M^2\biggr)\varphi_{ext}(x,x_5)=0 \eqno(3.3b)$$
for the ``external'' regions  $II,IV$ \footnotemark.

\footnotetext{It must be noted, that in
the usual formulation of the 
conformal transformations in the coordinate space 
one can also divide  any field operator and
the corresponding equations of motions into two 5D 
parts using  inversion
 ${x_{\mu}}'=-\ell^2 x_{\mu}/x^2$. In particular, 
starting from the 6D invariant form $\xi_A\xi^A=0$ with
$x_{\mu}=\xi_{\mu}{\ell}/(\xi_5+\xi_6)$ we have
$x^{\mu}x_{\mu}=-\ell^2(\xi_5-\xi_6)/(\xi_5+\xi_6)$. This condition
can be projected into two 5D hyperboloid
$x^{\mu}x_{\mu}\pm x_5^2=\pm\ell^2$
with $x_5^2=2\xi_5(\ or\ \xi_6)\ell^2/(\xi_5+\xi_6)$.
Then we get the internal  and the external 5D regions with the boundary values 
$x^2=0,\pm \ell^2$ as it was done for 
$q^2,q_5^2$ variables in table 1.
For $\Phi(x)$ we can introduce an analogue to (3.2a,b)
boundary conditions
$\Phi(x)=\varphi_{inr}(x,x_5=t)+\varphi_{ext}(x,x_5=t)$.
In the such constructions 
the  operator \cite{Gatto} ${ \cal M}_{\mu\nu}(x)
=g_{\mu\nu}-2x_{\mu}x_{\nu}/x^2={\cal
M}_{\mu\nu}(1/x)$  has the properties of the metric tensor
${\cal M}_{\mu\nu}(x){\cal M}^{\nu\sigma}(x)=\delta_{\mu}^{\sigma}$, ${\cal
M}_{\mu\nu}(x)x^{\nu}=-x_{\mu}$ and $\partial/\partial
x^{\mu}=1/{x'}^2 {\cal M}_{\mu\nu}(x')\partial/\partial
{x'}_{\nu}$. In this approach one can simplify the
5D and 4D equations in the  
conformal field theory in the coordinate space.}

We introduce the boundary condition for
the operator ${i\over M}{ {\partial}/ { \partial
x_5}}\varphi_{ }(x,x_5)$ which  is generated by 
fifth dimension 

$${i\over M}{ {\partial}\over { \partial x_5}}\varphi_{ }(x,x_5) =
\eta_a\varphi_{ }(x,x_5)+l_{ }(x,x_5);\ \ \ a=1,2\equiv inr,ext, \eqno(3.4)$$

where
$$\eta_{inr}=\sqrt{|1-m^2/M^2|};\ \ \ \eta_{ext}=\sqrt{1+m^2/M^2}.\eqno(3.5)$$
Acting with  $M^2(i/M\partial/\partial x_5 +\eta_a)$ ¢
on the relation (3.4) we get

$$\Bigl({ {\partial^2}\over { \partial x_5 \partial x^5}} +M^2-m^2\Bigr)
\varphi_{inr}(x,x_5)=
-M^2\Bigl({i\over M} {{\partial}\over { \partial x_5}}+\eta_{inr}\Bigr)
l_{inr}(x,x_5)  \eqno(3.6a)$$
and
$$\Bigl({ {\partial^2}\over { \partial x_5 \partial x^5}} +M^2+m^2\Bigr)
\varphi_{ext}(x,x_5)= -M^2\Bigl({i\over M}{ {\partial}\over {
\partial x_5}}+\eta_{ext}\Bigr) l_{ext}(x,x_5).  \eqno(3.6b)$$
Combining these equations with (3.3a,b) we obtain
$$\Bigl({ {\partial^2}\over { \partial x_{\mu} \partial x^{\mu}}} +m^2\Bigr)
\varphi_{inr}(x,x_5)=
M^2\Bigl({i\over M}{ {\partial}\over { \partial x_5}}+\eta_{inr}\Bigr)
l_{inr}(x,x_5)$$
$$\equiv j_{inr}(x,x_5)\eqno(3.7a)$$

$$\Bigl({ {\partial^2}\over { \partial x_{\mu} \partial x^{\mu}}} +m^2\Bigr)
\varphi_{ext}(x,x_5)=
-M^2\Bigl({i\over M}{ {\partial}\over { \partial x_5}}+\eta_{ext}\Bigr)
l_{ext}(x,x_5)$$
$$\equiv j_{ext}(x,x_5),\eqno(3.7b)$$
where 
$j_{a}(x,x_5)$  are determined via $l_{a}(x,x_5)$.

Solutions of eq.(3.7a,b) determine the corresponding solution of the 4D
equation (3.1a) with¨
$$J(x)\equiv J(x,x_5=t_5)=j_{inr}(x,x_5=t_5)+j_{ext}(x,x_5=t_5)
\eqno(3.8)$$

It must be  noted, that one can rewrite
 the boundary condition (3.4) in the  
integral form

$$\varphi_{a}(x,x_5)=e^{-iM(x_5-t_5)}\biggl\{\varphi_a(x,x_5=t_5)-
{1\over {\eta_a}}l_a(x,x_5=t_5)\bigl[ e^{2iM\eta_a(x_5-t_5)} -1\bigr]$$
$$
-{1\over {2iM\eta_a}}\int_{t_5}^{x_5} dz_5 j_a(x,z_5) e^{-iM\eta_a(z_5-t_5)}
\bigl[ e^{2iM\eta_a(x_5-t_5)} -e^{2iM\eta_a(z_5-t_5)}\bigr]
\biggr\}. \eqno(3.9)$$
For noninteracting particles, when
  $l_{a}(x,x_5)=0$ and $j_{a}(x,x_5)=0$, equations (3.7a,b) and the
constraints (3.3a,b) and (3.4) coincide with the analoguous
equations and constraints from ref. \cite{K2,K3} 
with the invariant form
$q_{\mu}q^{\mu}+ q_5^2=M^2$ 
or $q_{\mu}q^{\mu}- q_5^2=- M^2$.
For the infinitely large scale parameter $M\to \infty$ 
this formulation \cite{K1,K2,K3}  transforms into  the usual
 4D quantum field theory with the non restricted mass spectrum.

{\underline {\bf Consistency condition for the 5D equation of motion
 (3.7a,b) and the }}
\newline
{\underline {\bf boundary conditions (3.3a,b) and (3.4).:}}\ \ \ \ \ 
Combining eq.(3.7a,b) and (3.6a,b) we find
$$\Bigl({ {\partial^2}\over { \partial x_{\mu} \partial x^{\mu}}}
{ {\partial^2}\over { \partial x_{5} \partial x^{5}}}-
{ {\partial^2}\over { \partial x_{5} \partial x^{5}}}
{ {\partial^2}\over { \partial x_{\mu} \partial x^{\mu}}}\Bigr)
\varphi_{inr,ext}(x,x_5)$$
$$=
\mp\Bigl({ {\partial^2}\over { \partial x_{\mu} \partial x^{\mu}}} \pm
{ {\partial^2}\over { \partial x_{5} \partial x^{5}}}\mp M^2
\Bigr)j_{a}(x,x_5)=0.\eqno(3.10a)$$

According to this relation, 
the 5D equations of motion (3.7a,b) 
are consistent with the boundary conditions (3.3a,b) and 
(3.4), if
$j_{a}(x,x_5)$ and $l_{a}(x,x_5)$ 
are embedded in hyperboloid  $q^2\pm q_5^2=\pm M^2$, i.e.
in analogue to $\varphi_{a}(x,x_5)$ the operators
$j_{a}(x,x_5)$ and $l_{a}(x,x_5)$ must satisfy the
conditions

$$\biggl({{\partial^2}\over{\partial x^{\mu}\partial x_{\mu}}}+
{{\partial^2}\over{\partial x^5\partial
x_5}}+M^2\biggr)j_{inr}(x,x_5)=0, \eqno(3.10b)$$

$$\biggl({{\partial^2}\over{\partial x^{\mu}\partial x_{\mu}}}-
{{\partial^2}\over{\partial x^5\partial
x_5}}-M^2\biggr)j_{ext}(x,x_5)=0. \eqno(3.10c)$$

Using the conditions (3.3a,b) $\varphi_{a}(x,x_5)$ may be represented
as

$$\varphi_{inr}(x,x_5)={2M\over{(2\pi)^4}}\int d^5q
e^{-iq_5x^5} \delta(q^2+q_5^2-M^2)
[\theta(q^2)\theta(M^2-q^2)+\theta(-q^2)\theta(-M^2-q^2)]$$
$$\Bigl[ e^{-iqx}\varphi^{(+)}_{inr}(q,q_5)+e^{iqx}\varphi^{(-)}_{inr}(q,q_5)
\Bigr],\eqno(3.11a)$$
and
$$\varphi_{ext}(x,x_5)={2M\over{(2\pi)^4}}\int d^5q
e^{-iq_5x^5} \delta(q^2-q_5^2+M^2)
[\theta(q^2)\theta(-M^2+q^2)+\theta(-q^2)\theta(M^2+q^2)]$$
$$\Bigl[ e^{-iqx}\varphi^{(+)}_{ext}(q,q_5)+e^{iqx}\varphi^{(-)}_{ext}(q,q_5)
\Bigr].\eqno(3.11b)$$


From (3.10b,c) we get the same representation for source operator

$$j_{inr,ext}(x,x_5)={2M\over{(2\pi)^4}}\int d^5q
e^{-iq_5x^5} \delta(q^2\pm q_5^2\mp M^2)
[\theta(q^2)\theta(\pm M^2 \mp q^2)+\theta(-q^2)\theta(\mp M^2\mp q^2)]$$
$$\Bigl[ e^{-iqx}j^{(+)}_{inr,ext}(q,q_5)+e^{iqx}j^{(-)}_{inr,ext}(q,q_5)
\Bigr].\eqno(3.12)$$

The same representation is valid  for $l_{inr,ext}(x,x_5)$.

\par
This formulation has a number of common properties with the  other
5D field-theoretical approaches based on the proper time 
method \cite{IZ,Fanchi,Land,Kubo,Green}, where
 $x_5^2\equiv\tau=x_0^2-{\bf x}^2\equiv x_{\mu}x^{\mu}$.
From this point of view the boundary conditions
(3.4) can be treated as an evolution equation.
On the other hand the fifth momentum $q_5$
 is singlevalued determined  via the scale parameter
 $\lambda$  (see table 2). Therefore   
 $x_5$ may be provided  with the  a scale interpretation if we take
$\lambda^{-1}=ln(Mx_5)$.
Unlike  other 5D approaches \cite{Fanchi,Snyder,Yang,Hamilton},
in the present formulation field operators and
the source operators are defined in the 5D hyperboloid.

\newpage


\centerline{\bf{4.\ 5D Lagrangian approach}} \vspace{0.5cm}

\par
The 5D operators
 $\varphi_{inr}(x,x_5)$ and $\varphi_{ext}(x,x_5)$ are independent because
they are defined
in the different domains of $q^2\equiv q_{\mu}q^{\mu}$ and $q_5^2$.
Therefore the sought 5D Lagrangian ${\cal L}\equiv {\cal L}(x,x_5)$ 
we will construct using the two sets of the independent fields  
$\varphi_{inr}(x,x_5)$ and $\varphi_{ext}(x,x_5)$.

 $${\cal L}={\cal L}_0+{\cal L}_{INT}+{\cal L}_c,\eqno(4.1a)$$
where ${\cal L}_0$ stands for the noninteracting part 

$${\cal L}_0=\sum_{a=1,2}\Bigl[
{{\partial\varphi_a(x,x_5)}\over{\partial x_{\mu}}}
{{\partial\varphi^{+}_a(x,x_5)}\over{\partial x^{\mu}}}-m^2
\varphi_a(x,x_5)\varphi^{+}_a(x,x_5)\Bigr],\eqno(4.1b)$$

$a=1,2\equiv inr,ext$, ${\cal L}_{INT}\equiv{\cal L}_{INT}
(\varphi_a,\varphi^{+}_a, {{\partial\varphi_a}/{\partial
x_{\mu}}}, {{\partial\varphi^{+}}_a/{\partial x^{\mu}}};
{{\partial\varphi_a}/{\partial x_{5}}},
{{\partial\varphi^{+}_a}/{\partial x^{5}}})$
is the interacting part of Lagrangian and
 ${\cal L}_c$ generates  the constraint (3.4)

$${\cal L}_c=M^2\sum_{a=1,2}
|{i\over M}{{\partial\varphi_a}\over{\partial x_{5}}}-\eta_{a}\varphi_a
-l_a(x,x_5)|^2.\eqno(4.1c)$$
where $l_a(x,x_5)\equiv l_a(\varphi_a,\varphi^{+}_a,
{{\partial\varphi_a}/{\partial x_{\mu}}},
{{\partial\varphi^{+}_a}/{\partial x^{\mu}}};
{{\partial\varphi_a}/{\partial x_{5}}},
{{\partial\varphi^{+}_a}/{\partial x^{5}}})$ 
is operator from the constraint (3.4).
\par
Next we consider   action
$${\cal S}(x_5)=\int d^4x {\cal L}(x,x_5)\eqno(4.2)$$
and its  variation 
under the conformal transformations (1.1a)-(1.1e)
$$\delta q_{\mu}=\delta h_{\mu}+\delta \Lambda_{\mu\nu}q^{\nu}+
\delta \lambda q_{\mu}+(q^2\delta {\hbar}_{\mu}- 2q^{\nu}\delta
{\hbar}_{\nu} q_{\mu})/M^2\eqno(4.3a)$$ 
where  $\delta h_{\mu}(\delta
\hbar_{\mu})$, $\delta \Lambda_{\mu\nu}q^{\nu}=-\delta
\Lambda_{\nu\mu}q^{\nu}$ ¨ $\delta \lambda$ 
stands for the infinitesimal parameters of the corresponding transformations. 

Translation $q_{\mu}'=q_{\mu}+h_{\mu}$ does not change
 $x_{\mu}$. Therefore the variation of coordinates
 $\delta x_{\mu}=x_{\mu}'-x_{\mu}$, generated by variation of four momenta
 $q_{\mu}$  (4.3), includes only the rotation and the scale transformations

$$\delta x_{\mu}=\delta \Lambda_{\mu\nu}^{-1}x^{\nu}-
\delta \lambda x_{\mu}\eqno(4.3b)$$

In the considered formulation
$x_5$ is independent variable. Therefore we take

$${ {\delta x_5}\over {\delta x_{\mu}} }=
\delta {{d x_5}\over {d x_{\mu}} }=0\eqno(4.4)$$
which is consistent with our choice of action (4.2).

Now we have
$$\delta {\cal S}(x_5)=\sum_{a=1,2}\biggl\{\int d^4 x\biggl[
{{\partial{\cal L}}\over{\partial \varphi^{+}_a(x,x_5)}}
-{{\partial}\over{\partial x_{\mu}}} \Bigl(
 {{\partial{\cal L}}\over{\partial
[{{\partial \varphi^{+}_a(x,x_5) }/{\partial x^{\mu} ] }} }}
    \Bigr)\biggr]{\overline \delta}\varphi_a^{+}(x,x_5)$$

$$+\int d^4 x {{d}\over{d x_{\mu}}} \Bigl[
{{\partial{\cal L}}\over{\partial \ [
{{\partial \varphi^{+}_a(x,x_5) }/{\partial x^{\mu} ] }} }}
{\overline  \delta}\varphi_a^{+}(x,x_5) + {\cal L}(x,x_5) \delta x^{\mu}
\Bigr]\biggr\}$$

$$+\int d^4 x  \Bigl[
{{\partial{\cal L}}\over{\partial  [
{{\partial \varphi^{+}_a(x,x_5) }/{\partial x^{5}   ] }} }}
 {\overline  \delta}\bigl(
{{\partial}\over{\partial x_{5}}}\varphi_a^{+}(x,x_5)\bigr) +
{{d{\cal L} }\over{d x_{5}}}\delta x^{5}
\Bigr]$$
$$\ +\ hermitian\ cojugate\eqno(4.5)$$
where  ${\overline  \delta}$ denotes a variation of form of the corresponding 
expression. Substituting
 ${{d{\cal L} }/{d x_{5}}}$ in (4.5)  we obtain
$$\delta {\cal S}(x_5)=\sum_{a=1,2}\biggl\{\int d^4 x\biggl[
{{\partial{\cal L}}\over{\partial \varphi^{+}_a(x,x_5)}}
-{{\partial}\over{\partial x_{\mu}}} \Bigl(
 {{\partial{\cal L}}\over{\partial
[{{\partial \varphi^{+}_a(x,x_5) }/{\partial x^{\mu} ] }} }}
    \Bigr)\biggr]\biggl[{\overline \delta}\varphi_a^{+}(x,x_5)+
{{\partial \varphi^{+}_a(x,x_5) }\over{\partial x^{5} }}\delta x_5
\biggr]\eqno(4.6a)$$
$$+\int d^4 x {{d}\over{d x_{\mu}}} \Bigl[
{{\partial{\cal L}}\over{\partial \ [
{{\partial \varphi^{+}_a(x,x_5) }/{\partial x^{\mu} ] }} }}
{\overline  \delta}\varphi_a^{+}(x,x_5) + {\cal L}(x,x_5) \delta x^{\mu}
\Bigr]\eqno(4.6b)$$
$$+\int d^4 x
{{\partial{\cal L}}\over{\partial  [
{{\partial \varphi^{+}_a(x,x_5) }/{\partial x^{5} ] }} }}
\Bigl[ {\overline  \delta}\bigl(
{{\partial}\over{\partial x_{5}}}\varphi_a^{+}(x,x_5)\bigr) +
{{\partial^2 \varphi^{+}_a(x,x_5) }\over{\partial {x_{5}}^2 }}\delta x_5
\Bigr]\eqno(4.6c)$$
$$+\int d^4 x {{d}\over{d x_{\mu}}}\Bigl[
{{\partial{\cal L}}\over{\partial \ [ {{\partial
\varphi^{+}_a(x,x_5) }/{\partial x^5 ] }} }} {{\partial
\varphi_a^{+}(x,x_5)}\over{\partial x_{\mu} }}\delta x_5
\Bigr]\biggr\}\eqno(4.6d)$$
$$\ +\ hermitian\ cojugate$$

In order to get $\delta {\cal S}(x_5)=0$  we will suppose
that every term of eq.(4.6) vanishes. Then 
for the every term separately we obtain the following equations:

\begin{itemize}

\item[$1.$]
The first term (4.6a) represents the equation of motion for
$\varphi_a^{+}(x,x_5)$ and $\varphi_a^{+}(x,x_5)$
$${{\partial{\cal L}}\over{\partial \varphi^{+}_a(x,x_5)}}
={{d}\over{d x_{\mu}}} \Bigl(
 {{\partial{\cal L}}\over{\partial
[{{\partial \varphi^{+}_a(x,x_5) }/{\partial x^{\mu} ] }} }}\Bigr);\ \ \
{{\partial{\cal L}}\over{\partial \varphi_a(x,x_5)}}
={{d}\over{d x_{\mu}}} \Bigl(
 {{\partial{\cal L}}\over{\partial
[{{\partial \varphi_a(x,x_5) }/{\partial x^{\mu} ] }}
}}\Bigr),\eqno(4.7a)$$

or
$$\Bigl({{\partial^2}\over{\partial x_{\mu} \partial x^{\mu}}}
+m_0^2\Bigr)\varphi_a(x,x_5)= {{\partial{\cal
L}_{INT}}\over{\partial \varphi^{+}_a(x,x_5)}} -{{d}\over{d
x_{\mu}}} \Bigl(
 {{\partial{\cal L}_{INT}}\over{\partial
[{{\partial \varphi^{+}_a(x,x_5) }/{\partial x^{\mu}] }} }}
    \Bigr)\equiv j_a(x,x_5)\eqno(4.7b)$$
which coincides with (3.7a,b).

\item[$2.$]
The next term (4.6b) relates to the 4D current conservation condition

$${\cal J}^{\mu}(x)=\sum_{a=1,2}{\cal J}^{\mu}_a(x,x_5=t_5),\eqno(4.11a)$$
where
$${\cal J}^{\mu}_a(x,x_5)={{\partial{\cal L}}\over{\partial \ [
{{\partial \varphi^{+}_a(x,x_5) }/{\partial x_{\mu} ] }} }}
{\overline  \delta}\varphi_a^{+}(x,x_5) + {\cal L}(x,x_5) \delta
x^{\mu}$$
$$\ +\ hermitian\ cojugate.\eqno(4.7b)$$

\item[$3.$]
Third term  (4.6c) contains
 $\partial\varphi_a(x,x_5) /\partial x^{5}$. This field may be treated
as independent due to fifth degrees of freedom \cite{K2,K3}.
Therefore we can introduce  a new kind of fields
$$\chi_a(x,x_5)={i\over M}
{{\partial\varphi_a(x,x_5) }\over{\partial x_{5} }},\eqno(4.8)$$
 ${\cal L}_0$ does not contain these fields and they are defined
via constraint Lagrangian ${\cal L}_c$. 
Using the variation principle and  the
independence of the fields $\chi_a(x,x_5)$ we get 
$$
{{\partial{\cal L}}\over{\partial  [ {{\partial
\varphi^{+}_a(x,x_5) }/{\partial x^{5}   ] }} }} =
{{\partial{\cal
L}}\over{\partial  [ {{\partial \varphi_a(x,x_5) }/{\partial x^{5}
] }} }}=
{{\partial{\cal
L}_c}\over{\partial  [ {{\partial \varphi_a(x,x_5) }/{\partial x^{5}
] }} }}=0 \eqno(4.9)$$ 
which implies

$$\chi_a-\eta_{a}\varphi_a-l_a(x,x_5)=-1/M^2
{{\partial{\cal L}_{INT} }\over{\partial \chi^{+}_a(x,x_{5})}}$$
$$+{{\partial l^{+}_a }\over{\partial \chi^{+}_a(x,x_{5})}}
\Bigl( \chi_a-\eta_{a}\varphi_a-l_a(x,x_5)\Bigr) +{{\partial l_a
}\over{\partial \chi^{+}_a(x,x_{5})}} \Bigl(
\chi^{+}_a-\eta_{a}\varphi^{+}_a-l_a(x,x_5)^{+}\Bigr).
\eqno(4.10)$$

Afterwards we restrict our formulation
with such 
${\cal L}_{INT}$ which are independent on  ${{\partial\varphi_a}/{\partial
x_{5}}}$ and  ${{\partial\varphi^{+}}/{\partial x^{5}}}$.
Then instead of (4.10) we get

$$\chi_a(x,x_5)-\eta_{a}\varphi_a(x,x_5)-l_a(x,x_5)=0,\eqno(4.11)$$
which coincides with  (3.4).

 Combining (3.2a,b) and (4.11) we get the connections between
$l_a(x,x_5)$ and $j_a(x,x_5)$

$$j_{a}(x,x_5)=(-1)^{a-1}
M^2\Bigl({i\over M}{ {\partial}\over { \partial
x_5}}+\eta_{a}\Bigr) l_{a}(x,x_5).\eqno(4.12)$$
which was presented in eq. (3.7a,b).

\item[$4.$]
The fourth term  (4.6d) contains the current operator
$$ J^{\mu}_a(x,x_5)=
{{\partial{\cal L}}\over{\partial \ [ {{\partial
\varphi^{+}_a(x,x_5) }/{\partial x^{\mu} ] }} }} {{\partial
\varphi_a^{+}(x,x_5)}\over{\partial x_{5} }} \delta x_5$$
$$\ +\ hermitian\ cojugate\eqno(4.13).$$

in the asymptotic region, where $j_a=0$ and $l_a=0$,
 $ J^{\mu}_a(x,x_5)$ has the same form as the electro-magnetic 
current operator

$$ J^{\mu}_a(x,x_5)=-i\eta_a\delta x_5\biggl\{
{{\partial{\cal L}_0}\over{\partial \ [
{{\partial \varphi_a(x,x_5) }/{\partial x^{\mu} ] }} }}
 \varphi_a(x,x_5)$$
$$-
{{\partial{\cal L}_0}\over{\partial \ [
{{\partial \varphi^{+}_a(x,x_5) }/{\partial x^{\mu} ] }} }}
 \varphi_a^{+}(x,x_5)
\biggr\}\eqno(4.14)$$

 and $ {{d}/{d x_{\mu}}}J^{\mu}_a(x,x_5)=0$.
For the interacting fields expression (4.13) vanishes if we require 
that $\delta x_5=0.$

\end{itemize}

Thus we have derived the equation of motion (3.7a,b), constraint
(3.4) and the expression for the conserved currents (4.7b) and (4.13)
using the variation principle. Combining these equations one can verify, that
$d {\cal S}(x_5) / dx_5=0,$ i.e. ${\cal S}(x_5)$ (4.2) is not
dependent on $x_5$. In particular, using the 5D equations of motion
(4.7a,b) we get $d {\cal S}(x_5) /d x_5
=\int d^4x d/d x_{\mu} \Bigl\{ \partial {\cal L}
/\partial[{{\partial \varphi^{+}_a(x,x_5)} /{\partial x^{\mu} }}]
\partial \varphi_a^+(x,x_5)/\partial x_{5}+\ í.\ á.\Bigr\}$
which vanishes according to fifth  current conservation condition 
(4.13).

\vspace{0.75cm}

\centerline{\bf{5.\ Construction of 5D  Lagrangian 
  ${\cal L}_{INT}$  via  $l_a(x,x_5)$ (3.4)}}
\vspace{0.75cm}

\par
In equations (3.7a,b) and  (3.4) 
operators $\varphi_a(x,x_5)$, $j_a(x,x_5)$ and $l_a(x,x_5)$ are 
defined on the shell of the hyperboloid  $q^2\pm q_5^2=\pm M^2$.
But the product of $\varphi_a(x,x_5)$ is not on  
$q^2\pm q_5^2=\pm M^2$ shell. Therefore in order to construct
some explicit  representation of eq.(3.7a,b) and (3.4)
it is convenient to find the 
simple off 
shell extension of $\varphi_a(x,x_5)$  
and the corresponding Lagrangians.
The off   $q^2\pm q_5^2\mp M^2$ shell operator
 ${\widetilde \varphi_a}(x,x_5)$ can be introduced as follows

$$\varphi_{a}(x,x_5)=\int d^5y {\widetilde \varphi_a}(x-y,x_5-y_5)
D_{a}(y,y_5),\eqno(5.1)$$
where

$${\widetilde \varphi_a}(x,x_5)={2M\over{(2\pi)^4}}\int d^5q
e^{-iqx-iq_5x^5}
[\theta(q^2)\theta(\pm M^2\mp q^2)+\theta(-q^2)\theta(\mp M^2\mp q^2)]$$
$$\Bigl[\varphi^{(+)}_{inr}(q,q_5)+\varphi^{(-)}_{inr}(-q,q_5)
\Bigr],\eqno(5.2a)$$
and

$$D_a(x,x_5)={1\over{(2\pi)^5}}\int d^5q
e^{-iqx-iq_5x^5}\delta(q^2\pm q_5^2\mp M^2).\eqno(5.3)$$
Substituting (5.2a) and (5.3) into (5.1) we obtain expressions (3.11a,b)  
after Fourier transforms.

Certainly,  
${\widetilde \varphi_a}(x,x_5)$ 
is not  determined via $\varphi_a(x,x_5)$ singlevalued. 
In particular,  we can take a scale-invariant representation
 
$${\widetilde \varphi_{a}}(x,x_5)={1\over{(2\pi)^4}M}\int d^5q (Q_a)^{-4}
e^{-iqx-iq_5x^5}$$
$$[\theta(q^2/(Q_a)^2)\theta(\pm M^2\mp q^2/(Q_a)^2)+\theta(-q^2/(Q_a)^2)
\theta(\mp M^2\mp q^2/(Q_a)^2)]$$
$$\Bigl[ \varphi^{(+)}_{a}(q/Q_a,q_5/Q_a)
+\varphi^{(-)}_{a}(-q/Q_a,q_5/Q_a)\Bigr],\eqno(5.2b)$$

where $(Q_a)^2=(q^2\pm q_5^2)/M^2$ and $(Q_a)^2=1$
on the surfaces $q^2\pm q_5^2=\pm M^2$. Then we obtain
an equivalent representation   
 of $\varphi_{a}(x,x_5)$ (3.11a,b)

$$\varphi_{a}(x,x_5)={2M\over{(2\pi)^4}}\int d^5q
e^{-iqx-iq_5x^5} \delta(q^2\pm q_5^2\mp M^2) (Q_a)^{-4}$$
$$[\theta(q^2/(Q_a)^2)\theta(\pm M^2\mp q^2/(Q_a)^2)+\theta(-q^2/(Q_a)^2)
\theta(\mp M^2\mp q^2/(Q_a)^2)]$$
$$\Bigl[ \varphi^{(+)}_{a}(q/Q_a,q_5/Q_a)
+\varphi^{(-)}_{a}(-q/Q_a,q_5/Q_a)\Bigr].\eqno(5.2c)$$

Using (5.2b) and (5.2c) we obtain the inverse to (5.1) 
representation

$${\widetilde \varphi_a}(x,x_5)=\int_{0^+}^{\infty} d\alpha
  \varphi_a(\alpha x,\alpha x_5)\eqno(5.4a)$$
which implies that
$$
{{\delta{\widetilde \varphi_a}(x,x_5)}\over {\delta\varphi_{a}(x,x_5)}}=
{{\delta{\widetilde \varphi_a}(x,x_5)}
\over {\delta\varphi_{a}(\beta x,\beta x_5)}}
\Bigr]_{_{\beta =1}} =1.\eqno(5.4b)$$

An  other operator
${\cal O}(x,x_5)\equiv \chi_a(x,x_5), l_a(x,x_5), j_a(x,x_5),
({\cal L}_a)_{INT}(x,x_5)$ may be determined
 via the corresponding off shell operator
${\widetilde {\cal O}}_a(x,x_5)$ in the same way

$${\cal O}_a(x,x_5)=\int d^5y {\widetilde {\cal O}_a}(x-y,x_5-y_5)
D_{a}(y,y_5)\eqno(5.5a)$$

and vice versa, using the 
  representation (5.2a,c) we get

$${\widetilde {\cal O}_a}(x,x_5)=\int_{0^+}^{\infty} d\alpha
  {\cal O}_a(\alpha x,\alpha x_5).\eqno(5.5b)$$

\par
The straightforward generalization of 
equations (3.7a,b) is not available 
for ${\widetilde \varphi_a}(x,x_5)$ (5.2b) and (5.4a).  
In particular, if we take in eq.(3.7a,b) 
$x_{\mu}=\alpha y_{\mu}$ and $x_{5}=\alpha y_{5}$, then after
  integration over $\alpha$ we get
$${ {\partial^2}\over{ \partial y_{\mu}\partial y^{\mu}}}
\int_{0^+}^{\infty}{{d\alpha}\over{\alpha^2}}\varphi_{a}(\alpha
y,\alpha y_5) +m^2{\widetilde \varphi_a}(y,y_5)= {\widetilde
j_a}(y,y_5).\eqno(5.6)$$

On the other hand starting from the equations of motion
$$\biggl[ { {\partial^2}\over{ \partial x_{\mu}\partial x^{\mu}}}
 +m^2\biggr]{\widetilde \varphi_a}(x-y,x_5-y_5)= {\widetilde
j_a}(x-y,x_5-y_5),\eqno(5.7)$$
we obtain the equations of motion (3.7a,b)
using integration over $y,y_5$ variables according to eq. (5.1). 
Certainly, in eq.(5.6) and in 
eq.(5.7) the different expressions of ${\widetilde \varphi_a}$ are assumed.

In the same way as eq.(4.7a,b) with the constraint (4.11)
we can  derive equation of motion (5.7) and the off shell constraint
 for the off $q^2\pm q_5^2\mp M^2$ shell Lagrangian

$${\widetilde{\cal L}_a}=({\widetilde{\cal L}_a})_0+
({\widetilde{\cal L}_a})_{INT}+({\widetilde{\cal L}_a})_c,\eqno(5.8a)$$
where $({\widetilde{\cal L}_a})_0$ stands for the noninteracting part,
$({\widetilde{\cal L}_a})_{INT}$ is  the interaction part
and $({\widetilde{\cal L}_a})_c$  generate the constraint

$${\widetilde\chi_a}(x,x_5)-\eta_{a}{\widetilde\varphi_a}(x,x_5)-
{\widetilde l_a}(x,x_5)=0\eqno(5.9a)$$
and have the form
$$({\widetilde{\cal L}})_c=M^2\sum_{a=1,2}
|{i\over M}{{\partial{\widetilde\varphi}_a}\over{\partial x_{5}}}
-\eta_{a}{\widetilde\varphi}_a
-{\widetilde l}_a(x,x_5)|^2.\eqno(5.9b)$$
Thus relation (5.1) enables us to obtain the straightforward off shell
representations of eq.(3.7a,b) and the constraint (3.4) using the same
as (4.1a,b,c) Lagrangian but with off 
$q^2\pm q_5^2\mp M^2$ shell operators ${\widetilde\varphi}_a$.
Now we consider some example of
 ${\widetilde l_a}(x,x_5)$ and the corresponding interaction Lagrangians:

\vspace{0.7cm}
\par
{\underline {\bf $\varphi^4$ model:}} \
\ \ \ \ 
The simplest 
 ${\widetilde l_a}$ which does not dependent on
 ${\widetilde\chi_a}(x,x_5)\equiv
i/M\ \partial {\widetilde\varphi_a} (x,x_5)/\partial x_5$ is
$${\widetilde l_a}=g_a{\widetilde\varphi_a}^2,\eqno(5.10)$$

Using the constraint (5.9a)  we get

$${i\over M}{{\partial{\widetilde\varphi_a}(x,x_5) }\over{\partial x_{5} }}
=\eta_{a}\
{\widetilde\varphi_a}(x,x_5)+{g_a}{\widetilde\varphi_a}^2.\eqno(5.11)$$

$${\widetilde j_{a}}(x,x_5)=\partial {\widetilde{\cal L}_{INT}}/\partial
{\widetilde \varphi^{+}_{a}}(x,x_5)
=(-1)^{a-1}M^2\Bigl({i\over M}{{\partial}\over{\partial x_{5} }}
+\eta_{a} \Bigr){\widetilde l_{a}}(x,x_5)$$
$$=(-1)^{a-1}M^2\Bigl(3g_a  \eta_{a}{\widetilde\varphi_a}^2 +
2g_a^2{\widetilde\varphi}^3\Bigr).\eqno(5.12)$$ 
The corresponding equation of motion can be derived using the following
Lagrangians

$${\widetilde{\cal L}}={1\over 2}\sum_{a=1,2}\Bigl[
{{\partial {\widetilde\varphi_a}}\over{\partial x_{\mu}}}
{{\partial{\widetilde\varphi_a}}\over{\partial x^{\mu}}}-m^2_a
{\widetilde\varphi_a}^2\Bigr]+ M^2\sum_{a=1,2} |\ {\widetilde\chi_a}-\eta_{a}
{\widetilde\varphi_a} -{g_a}{\widetilde\varphi_a}^2\ |^2+ 
{\widetilde {\cal L}}_{INT},\eqno(5.13)$$ 
where 

$$({\widetilde{\cal L}_a})_{INT}(x,x_5)
=(-1)^{a-1}M^2\Bigl(  {g_a} \eta_{a} {\widetilde\varphi}_a^3 +
{{g_a^2}\over 2} {{\widetilde \varphi}_a^4}\Bigr)\eqno(5.14)$$

Lagrangian  (5.13) has  the following attractive properties
\begin{itemize}

\item[${\bf I.}$]
The considered model is renormalizable, because
${\widetilde{\cal L}}_{INT}$ and ${\widetilde{\cal L}}_{á}$ contains  
${\widetilde \varphi}_a$ in the third and in the
fourth power.

\item[${\bf II.}$]

$({\widetilde{\cal L}_{inr}})_{INT}$ $(a=1)$ and 
$({\widetilde{\cal L}_{ext}})_{INT}$ $(a=2)$ have the opposite sign.

\item[${\bf III.}$]
The Lagrangian  (5.14) has a local minimum at
 $-2\eta_{a}/ g_a$ and a local maximum at  $-{3\over2}\eta_{a}/ g_a$.

\end{itemize}

\vspace{0.7cm}
\par
{\underline {\bf
Nonlinear $\sigma$ model:}} \ \ \ \ \
In this case we have pi-meson fields $\pi^{\pm}$, $\pi^{0}$
instead of $\varphi$. We choose   $l_a$ 
depending on the auxiliary fields   $\chi$ 
$${\widetilde l_a}^{\alpha}={1\over {4f_{\pi}^2}}({\widetilde\chi_a}^{\gamma}
 {{\widetilde\chi}^{\gamma}_a})
{\widetilde\pi}_a^{\alpha}\equiv {1\over
{4f_{\pi}^2}}{\widetilde\chi}^2
{\widetilde\pi}_a^{\alpha},\eqno(5.15)$$ 
where we have used well known isospin redefinition of the pi-meson fields
$\pi^{\pm}\equiv 1/2(\pi^1\pm i \pi^2);\ \ \ \pi^{0}\equiv \pi^3$;
$\alpha,\beta,\gamma=1,2,3$, $f_{\pi}=93MeV$ is the pi-meson decay
 constant and Lagrangian   
  is  choosing in the form
$${\widetilde{\cal L}}=\sum_{a=1,2}\biggl( {1\over 2}
{{\partial}\over{\partial{x_\mu}}} {\widetilde\pi}_a^{\alpha}
{{\partial}\over{\partial{x_\mu}}} {{\widetilde\pi}^{\alpha}_a}
+M^2\Bigl[
{\widetilde\chi_a}^{\alpha}-{\widetilde\pi_a}^{\alpha}-
{1\over {4f_{\pi}^2}}{\widetilde\chi_a}^2
{\widetilde\pi_a}^{\alpha}
\Bigr]^2\biggr) +{\widetilde{\cal L}}_{chir}+
{\widetilde{\cal L}}_{INT},\eqno(5.16)$$

where the second term generates the constraint between the
auxiliary field
  ${\widetilde\chi_a}^{\alpha}(x,x_5)=
{i/ M}{{\partial{\widetilde\pi_a}^{\alpha}(x,x_5) }/{\partial x_{5}}}$ 
and the $\pi$ meson field ${\widetilde \pi_a}^{\alpha}$

$${\widetilde\chi_a}^{\alpha}-{\widetilde\pi_a}^{\alpha}-
{1\over {4f_{\pi}^2}}{\widetilde\chi_a}^2
{\widetilde\pi_a}^{\alpha}=0.\eqno(5.17a)$$ 

This constraint coincides with the relation between the 
$\pi$ meson field and the
interpolating field in the nonlinear
$\sigma$-model \cite{Wei,Alf}
$${\widetilde\pi_a}^{\alpha}={1\over{1+
{ {{\widetilde\chi_a}^2 }\over{4f_{\pi}^2} } } } 
{\widetilde \chi_a}^{\alpha},\eqno(5.17b)$$

Third term of (5.16)  ${\widetilde{\cal L}}_{chir}$ reproduces the  
constraint between pi-meson fields and the auxiliary 
 $\sigma$-meson fields
$${\widetilde\pi_a}^2+{\widetilde\sigma}_a^2=f_{\pi}^2
\eqno(5.18)$$

and correspondingly

$$({\widetilde{\cal L}}_a)_{chir}(x,x_5)=
\Bigl({\widetilde\pi_a}^2+{\widetilde\sigma}_a^2-f_{\pi}^2\Bigr)^2.
\eqno(5.19)$$

In the usual  $\sigma$ model
the chiral symmetry is weakly broken with the additional Lagrangian
${\cal L}'=-f_{\pi}m_{\pi}^2\sigma$. In the considered model
the chiral symmetry breaking terms arise in  
${\widetilde{\cal L}}_{INT}$. This Lagrangian may be constructed
using the source operator ${\widetilde j_{a}}^{\alpha}$  
which is defined via operator (5.15)

$${\widetilde j_{a}}^{\alpha}(x,x_5)=
\partial ({\widetilde{\cal L}_a})_{INT}/
\partial {\widetilde\pi^{\alpha}}_{a}(x,x_5)
=(-1)^{a-1}M^2\Bigl({i\over M}{{\partial}\over{\partial x_{5} }}+1\Bigr)
{\widetilde l_{a}}^{\alpha}(x,x_5)$$

$$=(-1)^{a-1} M^2{{(f_{\pi}+{\widetilde\sigma_a})}
\over{{\widetilde\sigma_a}} } \Bigl[ 1+ f_{\pi}
{{(3f_{\pi}-{\widetilde\sigma_a})}
\over{(f_{\pi}-{\widetilde\sigma_a})^2} } \Bigr]
{\widetilde\pi_a}^{\alpha} .\eqno(5.20)$$ 

The corresponding Lagrangian is
$${\widetilde{\cal L}}_{INT}=-\sum_{a=1,2}(-1)^{a-1}M^2
\Bigl(f_{\pi}{\widetilde\sigma_a} +{1\over 2}{\widetilde\sigma_a}^2
+f_{\pi}  {{(f_{\pi}+{\widetilde\sigma_a})^2}
\over{(f_{\pi}-{\widetilde\sigma_a})} } \Bigr)
\eqno(5.21a)$$
which after expansion in  $\pi_a^2$ power series
takes the form
$${\widetilde{\cal L}}_{INT}=-\sum_{a=1,2}(-1)^{a-1}M^2
\Bigl(-{9\over 2}f_{\pi}^2+ 8{{f_{\pi}^4}\over{ {\widetilde
\pi_a}^2} }-{\widetilde \pi_a}^2 -{1\over{4f_{\pi}^2} }{\widetilde
\pi_a}^4-...\Bigr) \eqno(5.21b)$$
Thus  ${\widetilde{\cal L}}_{INT}$ for the  
$ext=a\equiv 2$   induces 
the real $\pi$ meson mass term if $M$ is fixed as
$$M={{m_{\pi}}\over{\sqrt{2} }}.\eqno(5.22)$$
and for the internal $\pi$ meson field ${\widetilde \pi}_{a=1}$
in (5.21b) appear only negative $m_{\pi}^2$, i.e. ${\widetilde\pi}_{a=1}$
remines to be massless.

The suggested 
5D Lagrangian allows us to reproduce exactly the nonlinear $\sigma$
 Lagrangian in the region with $q^2>M^2=m_{\pi}^2/2$. Moreover the
 considered model we have reproduced explicitely the
 chiral symmetry breaking term  in the $\sigma$ models 
$-m_{\pi}^2 f_{\pi}{\widetilde\sigma_a}$
in (5.21a) together with the 
other chiral symmetry breaking terms.
In the limit $m_{\pi}\to 0$ (i. e. $M^2\to 0$,) the 
above Lagrangian transforms into the free Lagrangian for the massless
pion. Note that the chiral symmetry breaking mechanism  
allowed us to fix the scale parameter of the conformal
transformation group via the pion mass.

\vspace{0.75cm}

\centerline{\bf{6.\ Models with the gauge transformations}}
\vspace{0.75cm}

\par
{\underline {\bf 
 Gauge transformation in the 4D and 5D coordinate space.}


Using the 6D rotations  
we can perform translations $q_{\mu}'=q_{\mu}-eA_{\mu}(q)$ 
(or $q_{\mu}'=q_{\mu}-eA_{\mu}(q,q_5)$) always supposing that
the 6D cone $\kappa_A\kappa^A=0$ and its 
  5D  projections $q^2\pm {q_5}^2=\pm M^2$ are invariant.
In particular, in analogue to  eq.(2.5), 
 translations $q_{\mu}'=q_{\mu}-eA_{\mu}(q)$ imply the following
transformations of the 6D variables
$$\kappa_{\mu}'=\kappa_{\mu}-ea_{\mu}(\kappa_A)\kappa_{+};\ \ \ \ \
 \kappa_{+}'=\kappa_{+};$$
$$\kappa_{-}'=\kappa_{-}-
e/M^2\Bigl(
a_{\nu}(\kappa_A)\kappa^{\nu}+\kappa^{\nu}a_{\nu}(\kappa_A)\Bigr)+
e^2/M^2\kappa_+
a_{\nu}(\kappa_A)a^{\nu}(\kappa_A).$$
 After these transformations  we get
${q^2}'=q^2-
e\Bigl(A_{\nu}(q,q_5)q^{\nu}+q^{\nu}A_{\nu}(q,q_5)\Bigr)+
e^2A_{\nu}(q,q_5)A^{\nu}(q,q_5)$ and
${q^2_5}'=q^2_5\mp
e\Bigl(A_{\nu}(q,q_5)q^{\nu}+q^{\nu}A_{\nu}(q,q_5)\Bigr)+
e^2A_{\nu}(q,q_5)A^{\nu}(q,q_5)$, where 
the sign $-$ corresponds to
  $q^2+q_5^2=M^2$ and $+$ relates
to  $q^2-q_5^2=-M^2$.
$A_{\nu}(q,q_5)$ is constructed by
$a_{\nu}(\kappa_A)$ according to eq.(2.11). As a result of these gauge 
transformations ´we have ${q^2}'\pm{q_5^2}'={q^2}\pm{q_5^2}$. 

In order to derive
equations of motions using the gauge transformation 
in the  4D and 5D coordinate space,
we consider first the  off  $q^2\pm {q_5}^2=\pm M^2$ shell 
5D equations 

$$\Bigl({ {\partial^2}\over { \partial x_{\mu} \partial x^{\mu}}} +m^2\Bigr)
{\widetilde\varphi}_{a}(x,x_5)= {\widetilde j}_{a}(x,x_5),\eqno(6.1a)$$
where the source operator ${\widetilde j}_{a}(x,x_5)$ is defined via 
the 4D gauge transformations $q_{\mu}'=q_{\mu}-eA_{\mu}(q,q_5)$
$${\widetilde j}_{a}(x,x_5)=\Bigl(ie {{\partial}\over{\partial x_{\mu}
}}{\widetilde A}_{\mu}^a(x,x_5)+ie {\widetilde
A}_{\mu}^a(x,x_5){{\partial}\over{\partial
x_{\mu}}}-e^2{\widetilde A}_{\mu}^a(x,x_5){\widetilde A}^{\mu\ a}
(x,x_5) \Bigr){\widetilde \varphi}_a(x,x_5).\eqno(6.1b)$$ 

Using relations (5.1) and (5.5a)
we can embed eq.(6.1a,b) on  $q^2\pm {q_5}^2=\pm M^2$ shell 
$$\Bigl({ {\partial^2}\over { \partial x_{\mu} \partial x^{\mu}}} +m^2\Bigr)
{\varphi}_{a}(x,x_5)= { j}_{a}(x,x_5).\eqno(6.1c)$$

Next we can determine ${\widetilde l}_a(x,x_5)$ via $j_a(x,x_5)$

$${\widetilde l}_a(q,q_5)=(-1)^{a-1}{{{\widetilde j}_a(q,q_5)}
\over {M(q_5+M\eta_a)}}\eqno(6.2a)$$
and reproduce the corresponding constraint
$${i\over M}{ {\partial}\over { \partial x_5}}{\widetilde \varphi}_{a}(x,x_5) =
\eta_a{\widetilde \varphi}_{a}(x,x_5)+{\widetilde l}_{a}(x,x_5).
\eqno(6.2b)$$
Afterwards one can build corresponding 
off $q^2\pm {q_5}^2=\pm M^2$ shell Lagrangian

$${\widetilde {\cal L}}=\sum_{a=1,2}\Bigl[
{{\partial{\widetilde\varphi}_a(x,x_5)}\over{\partial x_{\mu}}}
{{\partial{\widetilde\varphi}^{+}_a(x,x_5)}\over{\partial
x^{\mu}}}-m^2
{\widetilde\varphi}_a(x,x_5){\widetilde\varphi}^{+}_a(x,x_5)+ M^2
|{i\over M}{{\partial{\widetilde\varphi}_a}\over{\partial
x_{5}}}-\eta_{a}{\widetilde\varphi}_a -{\widetilde
l}_a(x,x_5)|^2\Bigr]$$
$$+\sum_{a=1,2}\biggl[-ie{\widetilde\varphi}_a(x,x_5)
{{ \stackrel{\longleftrightarrow}{\partial}}\over{\partial x_{\mu}} }
{\widetilde \varphi}^{+}_a(x,x_5) {\widetilde
A}^{a\mu}(x,x_5) +e^2{\widetilde A}_{a\mu}(x,x_5){\widetilde
A}^{a\mu}(x,x_5){\widetilde
\varphi}_a(x,x_5){\widetilde\varphi}^{+}_a(x,x_5)
\biggr].\eqno(6.2d)$$

This Lagrangian reproduces  the equation of motion (6.1a) and the
constraint (6.2b) using the 4D gauge transformation 
$q_{\mu}'=q_{\mu}-e{({\widetilde A}_a)}_{\mu}(q,q_5)$
in the off $q^2\pm {q_5}^2=\pm M^2$ shell regions. 
On  $q^2\pm {q_5}^2=\pm M^2$ shell we obtain eq.(6.1c).

On the other hand the 4D gauge  transformations 
$q_{\mu}'=q_{\mu}-e{A  }_{\mu}(q)$ generates the ordinary 
 4D gauge equations

$$\Bigl( { {\partial^2}\over { \partial x_{\mu} \partial x^{\mu}} }
+m^2\Bigr)
\Phi(x)= J(x)=\Bigl(ie {{\partial}\over{\partial x_{\mu}}}A_{\mu}(x)
                   +ie A_{\mu}(x){{\partial}\over{\partial x_{\mu}}}
                   -e^2A_{\mu}(x)A^{\mu}(x) \Bigr)\Phi(x),\eqno(6.3a)$$ 

This equation can be divided into two equations using eq.(3.8) and eq.(3.2a)
$$J(x)= j_{inr}(x,t_5)+j_{ext}(x,t_5),\ \ \ \ \ 
\Phi(x)= {\phi}_{inr}(x,t_5)+{\phi}_{ext}(x,t_5)\eqno(6.3b)$$ 
with arbitrary $M$ and

$$\Bigl( { {\partial^2}\over { \partial x_{\mu} \partial x^{\mu}} }
+m^2\Bigr)
\phi_a(x,x_5)= j_a(x,x_5)
                   =\Bigl(ie {{\partial}\over{\partial x_{\mu}}}A_{\mu}(x)
                   +ie A_{\mu}(x){{\partial}\over{\partial x_{\mu}}}
                   -e^2A_{\mu}(x)A^{\mu}(x)
		   \Bigr)\phi_a(x,x_5).\eqno(6.3c)$$ 

The equations (6.3c) are the 5D representation of the 4D equations (6.3a),
 where  $x_5=t_5$ and the 4D gauge 
transformation is assumed. The source operator (6.1b) differs from the 
source operator (6.3b). Thus the 4D and the 5D gauge
 transformations generates the different 4D equations
(6.3a).

Using the exact form of  $j_a(x,x_5)$ (6.3c) we can construct 
$${ l}_a(q,q_5)=(-1)^{a-1}{{{ j}_a(q,q_5)}
\over {M(q_5+M\eta_a)}}\eqno(6.4a)$$
and reproduce the corresponding constraint
$${i\over M}{ {\partial}\over { \partial x_5}}{ \phi}_{a}(x,x_5) =
\eta_a{ \phi}_{a}(x,x_5)+{ l}_{a}(x,x_5)
\eqno(6.4c)$$
This enables us to construct the corresponding 
on $q^2\pm {q_5}^2=\pm M^2$ shell Lagrangian.

$${ {\cal L}}=\sum_{a=1,2}\Bigl[
{{\partial{\phi}_a(x,x_5)}\over{\partial x_{\mu}}}
{{\partial{\phi}^{+}_a(x,x_5)}\over{\partial
x^{\mu}}}-m^2
{\phi}_a(x,x_5){\phi}^{+}_a(x,x_5)+ M^2
|{i\over M}{{\partial{\phi}_a}\over{\partial
x_{5}}}-\eta_{a}{\phi}_a -{
l}_a(x,x_5)|^2\Bigr]$$
$$+\sum_{a=1,2}\biggl[-ie{\phi}_a(x,x_5)
{{ \stackrel{\longleftrightarrow}{\partial}}\over{\partial x_{\mu}} }
{\phi}^{+}_a(x,x_5) 
A^{\mu}(x) +e^2{ A}_{\mu}(x)
A^{\mu}(x)
\phi_a(x,x_5){\phi}^{+}_a(x,x_5)
\biggr].\eqno(6.4d)$$

This Lagrangian reproduces the 4D equation of motion (6.3a) with the
conditions (6.3b). But Lagrangian (6.4d) differs from Lagrangian
(6.2d) due to  $A_{\mu}$ and 
$j_{a}(x,x_5)$ operators.

In the same way we can redefine the equation of motion for the
fermion field operator

$$\Bigl( i\gamma_{\mu}{ {\partial}\over { \partial x_{\mu}} } -m_{el}\Bigr)
\Psi(x)= e\gamma_{\mu}A^{\mu}(x)\Psi(x)\equiv J(x) \eqno(6.5)$$ 
using the appropriate  5D equation 

$$\Bigl( i\gamma_{\mu}{ {\partial}\over { \partial x_{\mu}} } -m_{el}\Bigr)
{\widetilde\psi}_a(x,x_5)= 
{\widetilde j}_a(x,x_5),\eqno(6.6a)$$

where ${\widetilde j}_a(x,x_5)$ is constructed from $J(x)$ (6.5) and
generally it does not coincide with the pure 5D gauge source operator
$e\gamma_{\mu}{\widetilde
A}^{\mu}_a(x,x_5){\widetilde\psi}_a(x,x_5)$.  The fermion field
${\widetilde\psi}_a(x,x_5)$ satisfies the following 
constraints 

$$ \Bigl({i\over M}{{\partial}\over{\partial
x_5}}-\eta_a\Bigr){\widetilde \psi}_a(x,x_5)={\widetilde
l}_a(x,x_5),\eqno(6.6b)$$ £¤¥
$$(-1)^{a-1}M^2
\Bigl({i\over M}{{\partial}\over{\partial
x_5}}+\eta_a\Bigr){\widetilde l}_a(x,x_5)=-{\widetilde 
j}_a(x,x_5). \eqno(6.6c)$$

In the equations (6.1)-(6.6c) $M$ is a free parameter. 
If $M=0$ or $M\to\pm\infty$ we obtain
the usual 4D quantum field formulation without splitting into 
the ``inr''
and ``out'' regions in the momentum space.

\par
{\underline {\bf Gauge $SU(2)\times U(1)$
theory}} (see for example \cite{ChL}) can by formulated in the
5D form using the following   
off $q^2\pm {q_5}^2=\pm M^2$ shell Lagrangian

$${\widetilde {\cal L}}(x,x_5)=\sum_{a=1,2}
({\widetilde {\cal L}}_a)_V(x,x_5)+({\widetilde {\cal
L}}_a)_{sk}(x,x_5)+({\widetilde {\cal L}}_a)_F(x,x_5),
\eqno(6.7)$$
where $({\widetilde {\cal L}}_a)_V$ contains the vector Yang-Mills
 fields $({\widetilde A}_a)_{\mu}^{\alpha}(x,x_5)$ 
and Abelian fields $({\widetilde B}_a)_{\mu}(x,x_5)$

$$({\widetilde {\cal L}}_a)_V=
  -{1\over 4}(F_a)_{\mu\nu}^{\alpha}{(F_a)^{\mu\nu}}^{\alpha}
-{1\over 4}(G_a)_{\mu\nu}{(G_a)^{\mu\nu}}$$ $$ -M^2|{i\over
M}{{\partial}\over {\partial x_5}}({\widetilde
A}_a)^{\alpha}_{\mu}- ({\widetilde A}_a)^{\alpha}_{\mu}
-({\widetilde l}_a^A)^{\alpha}_{\mu}|^2 -M^2|{i\over
M}{{\partial}\over{\partial x_5}}({\widetilde B}_a)_{\mu}-
({\widetilde B}_a)_{\mu}- ({\widetilde l}_a^B)_{\mu}|^2,
\eqno(6.8a)$$ where
$$(F_a)_{\mu\nu}^{\alpha}=
{{\partial}\over{\partial{x^\mu}}}({\widetilde
A}_a)^{\alpha}_{\nu}-
{{\partial}\over{\partial{x^\nu}}}({\widetilde
A}_a)^{\alpha}_{\mu}+ g\varepsilon^{\alpha\beta\gamma}
({\widetilde A}_a)^{\beta}_{\mu} ({\widetilde
A}_a)^{\gamma}_{\nu},\eqno(6.8b)$$

$$(G_a)_{\mu\nu}=
{{\partial}\over{\partial{x^\mu}}}({\widetilde B}_a)_{\nu}-
{{\partial}\over{\partial{x^\nu}}}({\widetilde B}_a)_{\mu}.
\eqno(6.8c)$$

We define  $({\widetilde l}_a^A)^{\alpha}_{\mu}$,
$({\widetilde l}_a^B)_{\mu}$  via
 $({\widetilde j}_a^A)^{\alpha}_{\mu}$, $({\widetilde j}_a^B)_{\mu}$
in the same way as in eq.(6.2a,b).
The interacting parts of the gauge fields
$({\widetilde B}_a)_{\mu}$ and
$({\widetilde A}_a)^{\alpha}_{\mu}$ are contained in the
fermion and in the scalar terms  $( {\widetilde
{\cal L}}_a)_F$ and $({\widetilde {\cal L}}_{a})_{sk}$.

The scalar part of Lagrangian  (6.7) is

$$({{\widetilde {\cal L}}_a})_{sk}=\Bigl(
{\cal D}_a^{\mu}{\widetilde \Phi}_a\Bigr)^{\dag} \Bigl({\cal
D}_{a\mu}{\widetilde \Phi}_a\Bigr)- M^2| {i\over
M}{{\partial}\over{\partial x_5}}({\widetilde \Phi}_a)-
{\widetilde \Phi}_a-{\widetilde l}^{\Phi}_a|^2  +
{({\widetilde{\cal L}}_a)}_{INT},\eqno(6.9a)$$
where $({\widetilde{\cal L}}_a)_{INT}$ contains the self-interaction 
term of the scalar particle and

$${\cal D}_{a\mu}{\widetilde \Phi}_a=
\Bigl({{\partial}\over{\partial{x^\mu}}}- ig {{\tau^{\alpha}}\over
2}({\widetilde A}_a)^{\alpha}_{\mu} - ig'({\widetilde
B}_a)_{\mu}\Bigr){\widetilde \Phi}_a .\eqno(6.9b)$$

Unlike  the standard $SU(2)\times U(1)$ theory, we start from the
massless  $\Phi_a(x,x_5)$ field and the Higgs mechanism we will reproduce
using

$${\widetilde l}^{\Phi}_{a}(x,x_5)=
-{f\over M}{\widetilde \Phi}_{a}(x,x_5) \Bigl({\widetilde
\Phi}_{a}^{\ast}(x,x_5) {\widetilde
\Phi}_{a}(x,x_5)\Bigr)^{1/2}.\eqno(6.10)$$

In the unitary gauge

$$
{\widetilde \Phi}_{a}(x,x_5)={\cal U}\Bigl({\widetilde
 \zeta}_a(x,x_5)\Bigr)
\left( \begin{array}{c} 0  \\
 {\widetilde \phi}_a(x,x_5)\end{array} \right), \eqno(6.11a)$$
where

$${\cal U}\Bigl({\widetilde \zeta}_a(x,x_5)\Bigr)=
exp\Bigl(i{\widetilde \zeta}_a(x,x_5)/v\Bigr). \eqno(6.11b)$$

We will assume, that ${\widetilde \zeta}_a$ is independent on $x_5$

$${{\partial}\over{\partial{x_5}}}
{\widetilde \zeta}_a(x,x_5)=0.\eqno(6.12)$$

Afterwards ${\widetilde l}_a$ (6.10) takes a form
$${\widetilde l}^{\Phi}_{a}(x,x_5)=-{f\over M}{\cal U}\Bigl({\widetilde
\zeta}_a(x,x_5)\Bigr)
\left( \begin{array}{c} 0  \\
{{\widetilde \phi}_a}(x,x_5) \end{array}
\right)\sqrt{\Bigl({{\widetilde \phi}_a}(x,x_5)\Bigr)^2},
\eqno(6.13)$$
and for ${i\over M}{{\partial}/{\partial
x_5}}{\widetilde \phi}_a$ we get

$${i\over M}{{\partial}\over{\partial x_5}}({\widetilde \phi}_a)
={\widetilde \phi}_a-{f\over M}{\widetilde
\phi}_a\sqrt{({\widetilde \phi}_a)^2}.\eqno(6.14)$$
Next for ${\widetilde j}_a$ and for
$({\widetilde {\cal L}}_a)_{INT}$ we have

$${\widetilde j}_{a}=(-1)^{a-1}{\cal U}({\widetilde \zeta}_a)M^2
\left( \begin{array}{c} 0  \\
1 \end{array} \right) \Bigl( -3f M{\widetilde
\phi}_{a}\sqrt{({\widetilde \phi}_a)^2}+2f^2 {\widetilde
\phi}_a^3\Bigr) \eqno(6.15)$$

and

$$({\widetilde {\cal L}}_a)_{INT}=(-1)^{a-1}\Bigl( -fM{\widetilde
\phi}_a^2\sqrt{({\widetilde \phi}_a)^2}
 +{{f^2}\over 2}{\widetilde \phi}_a^4\Bigr)\eqno(6.16)$$

For the positive $f$  Lagrangian  ${\widetilde {\cal L}}_{a=1\equiv
  inr}$ is similar to the self-interaction potential  
$$V(\Phi)=-\mu^2\Phi^2+\lambda\Phi^4.\eqno(6.17)$$
In particular ${\widetilde {\cal L}}_{1}$ has zero at
${\widetilde \phi}_1=0$ and ${\widetilde \phi}_1= \pm {{2M}/ f}$
and  ${\widetilde {\cal L}}_{1}$ has minima at ${\widetilde
\phi}_1=\pm {{3M}/ {2f}}$. It is important to note, that
${\widetilde {\cal L}}_{a=2\equiv ext}=-{\widetilde {\cal L}}_{a=1\equiv inr}$.
Therefore in ${\widetilde {\cal L}}_{2}$ only the negative $m^2$
may be appear. For the 
spontaneous symmetry breaking case  we define

$$<0|\phi_{a}(x,x_5)|0>={1\over{2}}
\left( \begin{array}{c} 0  \\
{\sl v}_a\end{array} \right),\eqno(6.18)$$ where
 ${\sl v}_a={\sl v}$ for
$a=1\equiv inr$ and ${\sl v}_a=0$ for $a=2\equiv ext$.
Afterwards we get

$${\widetilde \phi}_{1}(x,x_5)={1\over2 }
\left( \begin{array}{c} 0  \\
{\sl v}+{\widetilde \phi}'(x,x_5)\end{array} \right);\ \ \ \ \
 {\widetilde \phi}_{2}(x,x_5)={1\over{2}}
\left( \begin{array}{c} 0  \\
{\widetilde \phi}'(x,x_5)\end{array} \right) .\eqno(6.19)$$
Therefore Lagrangian  (6.16) takes the form
$$({\widetilde {\cal L}}_a)_{INT}=(-1)^{a-1}\Bigl(
-{{fM}\over 4}({{\widetilde \phi}_{a}}'+{\sl v}_a )^2
        \sqrt{({{\widetilde \phi}_{a}}'+{\sl v}_a )^2} + {{f^2}\over
16}({\widetilde \phi}_{a}'+{\sl v}_a
 )^{4}\Bigr). \eqno(6.20)$$

Thus instead of the mass term in
the usual 4D self interacting potential (6.17)
(see for instance ch. 8 and 11 of \cite{ChL})
in the 5D Lagrangian  (6.20) arise the following terms 
 $$\mu^2 \Bigl({1\over{\sqrt{2} }}(\Phi'+v)\Bigr)^2\Longrightarrow
 {{fM}\over 4} \Bigl({{\widetilde \phi}'}_{inr}+{\sl v}\Bigr)^{2}
        \sqrt{\Bigl({{\widetilde \phi}'}_{inr}+{\sl v}\Bigr)^{2}}
-{{fM}\over 4}      ({{\widetilde \phi}'}_{ext} )^2
         \sqrt{({{{\widetilde \phi}'}_{ext}} )^2 } ,\eqno(6.21)$$
which determine the mass spectrum.

The effective Lagrangian (6.21) has minima at
 ${\widetilde \phi}_1=\pm {{3M}/ {f}}$. Therefore
  ${\widetilde {\cal L}}_{a=1}$ does not contain
the linear terms  in ${\widetilde \phi}_1$, 
 i.e.   ${\widetilde j}_1$ does not include the  constant terms.
If we take $v=\sqrt{2}{\sl v}=3\sqrt{2}M/f$,
then in the considered model all expressions
for the  fermion masses  were reproduced, i.e.
 $m_{k}=f_kv/\sqrt{2}$, $k=el,u,d$
and the  $W,Z$ meson masses remain be the same 
$m_Z=m_W/cos{\theta_W}$. But, for the Higgs boson mass
from Lagrangian (6.21) we get $m_{higgs}^2=9/8M^2$.
Thus the principal difference between the suggested 5D formulation
and the standard
$SU(2)\times U(1)$ gauge field theory
consist in the symmetry breaking terms (6.21).
The scale parameter $M$  is
defined via the mass of the Higgs boson and it 
indicates  the border  $q^2=\pm M^2$,
 where the interaction of the scalar fields change the sign.

\vspace{0.5cm}

\centerline{\bf{7.\ Conclusion}} \vspace{0.5cm}
\par

This paper is devoted to
the conformal transformations of the interacted quantum fields
in the momentum space. 
The key point of the present formulation is the invariance of the 
6D cone $\kappa_{\mu}\kappa^{\mu}+\kappa_5^2-\kappa_6^2=0$
(1.2) under a conformal transformation of any field operator.
Therefore the 5D forms  $q^2\pm q_5^2=\pm M^2$ (2.2a,b), 
arising via  projection
of this 6D cone into 4D momentum space 
$q_{\mu}={ {\kappa_{\mu}}\over{\kappa_{+} }};\ \ \ \
\kappa_{\pm}=(\kappa_{5}\pm\kappa_{6})/M;\ \ \ \
\mu=0,1,2,3$ (1.3), are also invariant.  
This invariance was taken into account by derivation 
of a 5D equation of motion and  the corresponding 
5D Lagrangians for a interacting massive particles.
The suggested system of the 5D equations of
motion contains separately
the one-dimensional equations over the fifth coordinate
$x_5$. These one dimensional equations have the
form of the evaluation equations if $x_5^2={x_o^2-{\bf x}^2}$.
and the evolution operator of these equations $l_a(x,x_5)$
is connected with the particle source operator
 $j_a(x,x_5)=(-1)^{a-1}M^2(i/M\partial/\partial
x_5+M)l_a(x,x_5)$.

Special attention was given to the  inversion of momenta
$q'_{\mu}=-M^2 q_{\mu}/q^{2}$. In particular, the whole 
definition area of $q^2$ and $q_5^2$ was divided into
four internal and external regions (see table 1 and 2) which
are connected via inversion.
The 4D field operator 
 $\Phi(x)$ was constructed via 5D operators $\varphi_{inr}(x,x_5)$
and $\varphi_{ext}(x,x_5)$ using the boundary condition
 $\Phi(x)=\varphi_{inr}(x,x_5=t_5)+\varphi_{ext}(x,x_5=t_5)$ (3.2a).

The important
parameter of the  considered 5D formulation
is  the scale constant $M$ which is necessary by inversion
$q'_{\mu}=-M^2 q_{\mu}/q^{2}$ and  by determination
of  momenta $q_{\mu}$ through the 6D variables  
$\kappa_{\mu},\kappa_{\pm}$.  In models, where
the   operators
$l_a(x,x_5)$ from the fifth dimension boundary condition  are determined via  
the source operators $j_a(x,x_5)$
(see for example (6.2a,b) and (6.4a))
$M$ is not fixed. 
In the 
theories with the spontaneous breaking  symmetry
$M$ is defined via the proper mass of these theories.
For instance, for the nonlinear $\sigma$-model
  $M=m_{\pi}/\sqrt{2}$ and 
  $m_{higgs}=3/2\sqrt{2}M$ in the  5D formulation of the 
standard $SU(2)\times U(1)$ theory.
In the present 5D  nonlinear $\sigma$-model
 the $\lambda \sigma$ and other
chiral symmetry broken terms are explicitely reproduced. In 
the domain $q^2\ge m_{\pi}^2/2$ this 5D model 
coincides with the Weinberg nonlinear $\sigma$ model 
\cite{Wei,Alf} and in the internal domain ($0\le q^2<m_{\pi}^2/2$) the
 $\pi$ meson field is massless. It must be noted, that in the
considered 5D formulation the interaction Lagrangians and the 
corresponding source operators change their sign at the border 
 $q^2=\pm M^2$.

\vspace{0.5cm}

I am sincerely grateful to V.G.Kadyshevsky for numerous constructive 
and fruitful discussions. 



\end{document}